\documentclass[aps,prd,superscriptaddress,showpacs,reprint]{revtex4-1}
\usepackage[utf8]{inputenc}
\usepackage{textcomp}
\usepackage[dvipsnames]{xcolor}
\usepackage{amssymb,amsfonts,amsmath}
\usepackage[sort&compress]{natbib}
\usepackage[english]{babel}
\usepackage{braket}
\usepackage{graphicx}
\usepackage{bm}
\usepackage{bbm}
\usepackage{braket}
\usepackage{dcolumn}
\usepackage{IEEEtrantools}
\usepackage{pstricks,pst-coil,pst-node,pst-circ,pst-plot,pst-3dplot,
pst-solides3d,pst-sigsys,pstricks-add,pst-eucl,pst-grad}
\usepackage[normalem]{ulem}
\allowdisplaybreaks[1]
\def\biggg#1{{\hbox{$\left#1\vbox to30.0pt{}\right.$}}}

\def\Biggg#1{{\hbox{$\left#1\vbox to45.0pt{}\right.$}}}

\DeclareMathOperator{\str}{str}
\DeclareMathOperator{\tr}{tr}
\usepackage[caption=false]{subfig}
\usepackage{dcolumn}
\usepackage{lmodern}
\usepackage{microtype}

\begin{document}

\title{Fluctuation-induced higher-derivative couplings\\
and infrared dynamics of the Quark-Meson-Diquark Model}

\author{Niklas Cichutek}
\email[]{cichutek@th.physik.uni-frankfurt.de}
\affiliation{Institut f\"ur Theoretische Physik, 
Johann Wolfgang Goethe-Universit\"at, 
Max-von-Laue-Stra\ss e 1, 
D-60438 Frankfurt am Main, Germany}

\author{Florian Divotgey}
\email[]{fdivotgey@th.physik.uni-frankfurt.de}
\affiliation{Institut f\"ur Theoretische Physik, 
Johann Wolfgang Goethe-Universit\"at, 
Max-von-Laue-Stra\ss e 1, 
D-60438 Frankfurt am Main, Germany}

\author{J\"urgen Eser}
\email[]{eser@th.physik.uni-frankfurt.de}
\affiliation{Institut f\"ur Theoretische Physik, 
Johann Wolfgang Goethe-Universit\"at, 
Max-von-Laue-Stra\ss e 1, 
D-60438 Frankfurt am Main, Germany}

\date{\today}

\begin{abstract}
	In a qualitative study, the low-energy properties of the $\text{SO}\!\left(6\right)$-symmetric 
	Quark-Meson-Diquark Model as an effective model for two-color Quantum Chromodynamics 
	are investigated within the Functional Renormalization Group (FRG) approach. In particular, 
	we compute the infrared scaling behavior of fluctuation-induced higher-derivative couplings 
	of the linear Quark-Meson-Diquark Model and map the resulting renormalized effective action 
	onto its nonlinear counterpart. The higher-derivative couplings of the nonlinear model, 
	which we identify as the low-energy couplings of the Quark-Meson-Diquark Model, are therefore
	entirely determined by the FRG flow of their linear equivalents. This grants full access 
	to their scaling behavior and provides insights into conceptual aspects of purely bosonic 
	effective models, as they are treated within the FRG. In this way, the presented work is 
	understood as an immediate extension of our recent advances in the $\text{SO}\!\left(4\right)$-symmetric 
	Quark-Meson Model beyond common FRG approximations.
\end{abstract}

\maketitle

\section{Introduction}
\label{sec:introduction}

In a series of recent publications \cite{Eser:2018jqo, Divotgey:2019xea, Eser:2019pvd}, 
we presented a low-energy analysis of the (two-flavor) Quark-Meson Model (QMM), which 
was treated as a (linear) effective model of the fundamental theory of the strong interaction, 
Quantum Chromodynamics (QCD). To be precise, we expanded the effective action of the QMM 
up to (and including) fourth order in its bosonic field variables as well as their 
respective space-time derivatives. We applied the Functional Renormalization Group (FRG) 
approach \cite{Wetterich:1992yh, Bonini:1992vh, Ellwanger:1993mw, Morris:1993qb, 
Berges:2000ew, Bagnuls:2000ae, Pawlowski:2005xe, Dupuis:2020fhh} in order to (qualitatively) 
determine the scaling behavior of all higher-derivative couplings within the aforementioned 
derivative expansion, the latter of which substantially exceeds commonly 
used truncation schemes. As a result of this analysis, the low-energy limit 
of the QMM naturally emerges from the FRG integration of quantum 
fluctuations by transforming the linear realization of the $\text{SO}\!\left(4\right)$-symmetry 
(among the bosonic fields) into its nonlinear counterpart \cite{Gursey:1959yy, 
GellMann:1960np, Schwinger:1967tc, Weinberg:1968de, Callan:1969sn, Coleman:1969sm, 
Meetz:1969as, Divotgey:2019xea}. According to the given expansion, the nonlinear realization 
encodes a complete set of (mostly momentum-dependent) pion self-interactions and thus defines the 
low-energy couplings of the model. Such calculations including higher-derivative interactions 
are generally very valuable as an (additional) consistency criterion for effective models.

An important consequence of the mapping between the (linear) QMM and its (nonlinear) 
low-energy limit is given by the fact that the low-energy couplings of the model are obtained 
as functions of the linear higher-derivative couplings. This provides access to their 
scaling behavior, which yields a suitable starting point for reliable estimates 
of the validity ranges of purely bosonic low-energy models within the FRG. Particularly, 
by decomposing the flow equations of these couplings into bosonic and fermionic components,
it turns out that the overall FRG scaling is almost entirely driven by the fermionic 
degrees of freedom (d.o.f.), even down to lowest momentum scales ($\lesssim 100\ \text{MeV}$). 
This finding is also confirmed by functional QCD calculations \cite{Braun:2014ata, 
Mitter:2014wpa, Cyrol:2017ewj, Paris-Lopez:2018vjc, Alkofer:2018guy, Fu:2019hdw}
and implies a late decoupling of the composite dynamics from the fundamental 
fluctuations, already characterizing an upper reliability bound for the pionic 
description.

The intention of the present work is to extend our considerations by various points:
\begin{itemize}
\item[(a)] We verify our qualitative findings in Refs.\ \cite{Eser:2018jqo, Divotgey:2019xea, 
Eser:2019pvd} by applying the developed formalism to another low-energy model that is as 
well subject to many investigations related to the strong interaction, namely, the 
$\text{SO}\!\left(6\right)$-symmetric Quark-Meson-Diquark Model (QMDM); the low-energy 
limit of this effective model with the corresponding low-energy couplings and 
the FRG scaling hierarchy of fluctuations will be (qualitatively) determined. The latter
goes hand in hand with the computation of appropriate transition scales from the linear 
to the nonlinear symmetry realization.

\item[(b)] Different truncation schemes within the QMDM are discussed in greater detail
and the effect of higher-derivative couplings on the numerical outcome will be elucidated.

\item[(c)] We study high-energy cutoff dependences of the low-energy couplings of the 
model (in an elementary setup); the couplings are thereby exclusively generated from quantum 
fluctuations and thus interpreted as model predictions.
\end{itemize}
In general, the QMDM serves as an effective description of two-color QCD ($\text{QC}_{2}\text{D}$),
which shares many important properties with QCD itself, such as a chirally broken mesonic
phase. Moreover, on the lattice, it does not suffer from a sign problem, which makes its 
phase diagram much more accessible than the one of three-color QCD \cite{Nakamura:1984uz,
Hands:1999md, Hands:2000ei, Hands:2001ee, Chandrasekharan:2006tz, Hands:2006ve, Hands:2011ye,
Hands:2012yy, Cotter:2012mb, Boz:2013rca, Scior:2015vra, Wilhelm:2019fvp, Boz:2019enj, 
Bornyakov:2020kyz} (and references therein). 
Due to these features, $\text{QC}_{2}\text{D}$ and also the QMDM have extensively been explored 
over the past decades, see e.g.\ Refs.\ \cite{Kogut:1999iv, Kondratyuk:1991hf, Rapp:1997zu, 
Kogut:2000ek, Splittorff:2000mm, Splittorff:2002xn, Dunne:2002vb, Dunne:2003ji, Ratti:2004ra, 
Brauner:2006dv, Sun:2007fc, Brauner:2009gu, Kanazawa:2009ks, Kanazawa:2009en, Andersen:2010vu, 
Zhang:2010kn, Strodthoff:2011tz, Imai:2012hr, vonSmekal:2012vx, Strodthoff:2013cua, Khan:2015puu,
Contant:2019lwf}.

This paper is organized as follows: Sec.\ \ref{sec:methods} introduces the methods 
and geometric concepts of the work, where Sec.\ \ref{sec:QMDM} and Sec.\ \ref{sec:FRG} 
focus on the foundations of the QMDM and the FRG, respectively. Afterwards,
Sec.\ \ref{sec:NGaction} discusses the principles of the nonlinear symmetry realization 
of the low-energy limit of the QMDM. The results are presented in Sec.\ \ref{sec:results}
(the generation of higher-derivative couplings is demonstrated in Sec.\ \ref{sec:linearmodel},
while the transition to the nonlinear picture is carried out in Sec.\ \ref{sec:nonlinearmodel}).
Finally, cutoff dependences of the low-energy couplings are studied in Sec.\ \ref{sec:cutoff}
and a concluding summary as well as a scientific outlook are given in Sec.\ \ref{sec:summary}.

\section{Methods}
\label{sec:methods}

\subsection{Two-color Quantum Chromodynamics and the linear Quark-Meson-Diquark Model}
\label{sec:QMDM}

Formally, $\text{QC}_{2}\text{D}$ is given by an $\text{SU}\!\left(N_{c}=2\right)$ 
gauge theory, which is minimally coupled to $N_{f} = 2$ massive Dirac-flavors of quarks 
($\psi$) in the fundamental representation of the gauge group. Even though this appears to be 
a rather slight deformation of QCD, the pseudoreality of the fermion (quark) representation 
introduces some radical changes w.r.t.\ accidental symmetries of the quark sector 
and their respective breaking schemes. As the latter crucially affect the low-energy 
dynamics of $\text{QC}_{2}\text{D}$ and, hence, also the construction of the QMDM, we want 
to briefly review these aspects in the following: $\text{QC}_{2}\text{D}$ possesses an 
``enlarged'' chiral symmetry \cite{Pauli:1957, Gursey:1958}, i.e., $\text{SU}\!\left(4\right) 
\cong \text{SO}\!\left(6\right)$, which is spontaneously broken down to its 
$\text{Sp}\!\left(4\right)\cong \text{SO}\!\left(5\right)$ subgroup due to an 
antisymmetric vacuum \cite{Strodthoff:2011tz, Strodthoff:2013cua, Khan:2015puu},
\begin{equation}
	\text{SO}\!\left(6\right) \longrightarrow \text{SO}\!\left(5\right) \! . \label{eq:SSB}
\end{equation}
This implies the occurrence of five pseudo-Nambu-Gold\-stone bosons (pNGBs) on the coset 
space $\text{SU}\!\left(4\right)\!/\text{Sp}\!\left(4\right)\cong\text{SO}\!\left(6\right)
\!/\text{SO}\!\left(5\right)$; three of them are given by the pions $\vec{\pi}$ 
and the remaining pNGBs are identified with the real components $\pi^{4,5}$ of two 
diquark fields. They manifest the meson-baryon symmetry, which is furnished by the 
flavor symmetry of $\text{QC}_{2}\text{D}$.

Those symmetry aspects are sufficient to construct the QMDM as a simple generalization 
of the QMM. To this end, we make use of the local isomorphisms stated above and assign 
the mesonic fields to an Euclidean field space vector $\varphi$,
\begin{equation}
	\varphi=\begin{pmatrix} \boldsymbol{\pi} \\ \sigma \end{pmatrix} \! ,
	\label{eq:phi}
\end{equation}
in which $\boldsymbol{\pi} = \left(\vec{\pi},\pi^{4},\pi^{5}\right)$ denotes the 
five-vector of pNGB fields and $\sigma$ is their chiral partner. The (Euclidean) 
vector $\varphi$ carries the fundamental representation of $\text{SO}\!\left(6\right)$.

The Lagrangian of the QMDM can then be written as
\begin{align}
	\mathcal{L}_{\text{QMDM}}&=\frac{1}{2}\bigl(\partial_{\mu} \varphi\bigr)\cdot\partial^{\mu}
	\varphi - V\!\left(\rho\right) + h_{\text{ESB}}\sigma \nonumber\\
	&\quad +\bar{\psi}\Bigl[i\gamma^{\mu}\partial_{\mu} - y\bigl(\sigma \tau_0 + i\gamma_{5}
	\vec{\pi}\cdot\vec{\tau}\bigr)\Bigr]\psi \nonumber \\ 
	&\quad +\frac{y}{2}\Bigl[\psi^{T}C\gamma_{5}i\sigma_{2}\tau_{2}\bigl(\pi^4-i\pi^5\bigr)
	\psi + \text{h.c.}\Bigr], \label{eq:QMDM}
\end{align}
where $\tau_{0} = \mathbbm{1}_{2}$ and $\vec{\tau}$ are the identity and the three Pauli 
matrices acting on flavor indices, respectively. Moreover, $C = -i\gamma_{2}\gamma_{0}$ 
symbolizes the charge conjugation operator and $\sigma_{2}$ corresponds to the second 
Pauli matrix in color space. The coupling $y$ is the Yukawa interaction.

In its tree-level approximation, the scalar potential $V$ is a quadratic polynomial 
in the $\text{SO}\!\left(6\right)$-invariant $\rho \equiv \varphi\cdot\varphi$,
\begin{equation}
	V\!\left(\rho\right) = \frac{m_{0}^{2}}{2}\rho + \frac{\lambda}{4}\rho^{2},
	\label{eq:TreePot}
\end{equation}
and therefore allows for a modeling of spontaneous chiral-symmetry breaking, 
cf.\ Eq.\ (\ref{eq:SSB}). To be precise, the breaking is realized by the introduction 
of an order parameter $\sigma_{0}$ that arises in the shape of a nonvanishing vacuum 
expectation value for the $\sigma$ field,
\begin{equation}
	\sigma \rightarrow \sigma + \sigma_{0}.
	\label{eq:vev}
\end{equation} 
In addition, the term $\sim h_{\text{ESB}}\sigma$ in Eq.\ (\ref{eq:QMDM}) describes 
the explicit breaking of chiral symmetry by tilting the (total) potential 
($V - h_{\mathrm{ESB}}\sigma$) into the direction of the scalar resonance $\sigma$.

\subsection{Effective action from the Functional Renormalization Group approach}
\label{sec:FRG}

Similar to Refs.\ \cite{Eser:2018jqo, Divotgey:2019xea, Eser:2019pvd}, we use the 
FRG to determine the scale evolution of the low-energy couplings of the QMDM. In
general, this method can be formulated in terms of a functional differential equation, 
the so-called Wetterich equation \cite{Wetterich:1992yh}, that describes the FRG 
flow of the effective average action $\Gamma_{\! k}$,
\begin{equation}
	\partial_{k}\Gamma_{\! k} = \frac{1}{2}\str\!\left[\left(\Gamma_{\! k}^{(2)} + 
	R_{k}\right)^{-1}\!\partial_{k}R_{k}\right] \! ,
	\label{eq:wetteq}
\end{equation}
where $k$ represents the FRG (energy-momentum) scale, $\smash{\Gamma_{\! k}^{(2)}}$ 
is the two-point function, and $R_{k}$ denotes an infrared (IR) regulator. The latter 
acts as a scale-dependent mass contribution for soft modes and implements the 
Wilsonian renormalization group idea of a momentum-shell-wise integration over 
the quantum fluctuations of the system \cite{Wilson:1970ag, Wilson:1971bg, 
Wilson:1971dh}. Hence, the effective average action $\Gamma_{\! k}$ is a 
functional that interpolates between the classical action, initialized at an 
ultraviolet (UV) cutoff $\Lambda_{\text{UV}}$, and the full effective (quantum) 
action $\Gamma_{\! k\, =\, 0} \equiv \Gamma$ in the IR.

The structure of Eq.\ (\ref{eq:wetteq}) implies an infinite tower of coupled 
differential equations. Thus our ability to solve the Wetterich equation depends 
on an appropriate truncation of correlation functions. In this light, we choose 
the following derivative expansion for the Euclidean effective average action:
\begin{align}
	\Gamma_{\! k} &= \int_{x} \biggl\{\frac{Z_{k}}{2}\bigl(\partial_{\mu}\varphi\bigr)\cdot 
	\partial_{\mu}\varphi + U_{k}\!\left(\rho\right) - h_{\text{ESB}}\sigma \nonumber \\[-0.1cm]
	&\qquad\;\;\;\;+ C_{2,k}\bigl(\varphi\cdot\partial_{\mu}\varphi\bigr)^{2} + Z_{2,k}\;
	\varphi^{2}\bigl(\partial_{\mu}\varphi\bigr)\cdot\partial_{\mu}\varphi \nonumber \\
	&\qquad\;\;\;\;- C_{3,k}\!\left[\bigl(\partial_{\mu}\varphi\bigr)\cdot\partial_{\mu}\varphi
	\right]^{2} - C_{4,k}\!\left[\bigl(\partial_{\mu}\varphi\bigr)\cdot\partial_{\nu}\varphi
	\right]^{2} \nonumber \\
	&\qquad\;\;\;\;- C_{5,k}\;\varphi\cdot\bigl(\partial_{\mu}\partial_{\mu}\varphi\bigr)\bigl(
	\partial_{\nu}\varphi\bigr)\cdot\partial_{\nu}\varphi \nonumber \\
	&\qquad\;\;\;\;- C_{6,k}\;\varphi^{2}\bigl(\partial_{\mu}\partial_{\nu}\varphi\bigr)\cdot
	\partial_{\mu}\partial_{\nu}\varphi \nonumber \\
	&\qquad\;\;\;\;- C_{7,k}\bigl(\varphi\cdot\partial_{\mu}\partial_{\nu}\varphi\bigr)^{2} - 
	C_{8,k}\;\varphi^{2}\bigl(\partial_{\mu}\partial_{\mu}\varphi\bigr)^{2} \nonumber \\
	&\qquad\;\;\;\;+ \bar{\psi}\Bigl[Z^{\psi}_{k}\gamma_{\mu}\partial_{\mu} + y_{k}\bigl(\sigma 
	\tau_0 + i\gamma_{5}\vec{\pi}\cdot\vec{\tau}\bigr)\Bigr]\psi \nonumber \\ 
	&\qquad\;\;\;\; +\frac{y_{k}}{2}\Bigl[\psi^{T}C\gamma_{5}i\sigma_{2}\tau_{2}\bigl(
	\pi^4-i\pi^5\bigr)\psi + \text{h.c.}\Bigr]\biggr\} . \label{eq:qmdmtrunc}
\end{align}
Here, we replaced Eq.\ (\ref{eq:TreePot}) by the effective potential $U_{k}\!\left(\rho\right)$ 
and established an obvious short-hand notation for the space-time integration. 
Furthermore, besides a scale-dependent Yukawa coupling $y_{k}$, we also introduced wave-function 
renormalization factors for both the bosonic and fermionic d.o.f., $Z_{k}$ and $Z^{\psi}_{k}$, 
respectively. Finally, as we are interested in the low-energy couplings of the QMDM, we 
supplemented the bosonic sector of Eq.\ (\ref{eq:QMDM}) by complete sets of higher-derivative 
couplings of $\mathcal{O}\!\left(\varphi^{4},\partial^{2}\right)$ and $\smash{\mathcal{O}\!\left(
\varphi^{4},\partial^{4}\right)}$. The truncation (\ref{eq:qmdmtrunc}) goes substantially 
beyond the commonly employed local potential approximation (LPA) (only the scale dependence 
of the effective potential $U_{k}$ is considered) or its minimal extension known as the
$\text{LPA}^{\prime}$ (involving wave-function renormalization). The chosen set of 
higher-derivative couplings conforms with the recently published fourth-order derivative 
expansion in the (generic) $\text{SO}\!\left(N\right)$-model \cite{DePolsi:2020pjk, 
Dupuis:2020fhh} [and Eq.\ (\ref{eq:qmdmtrunc}) even also includes fermionic d.o.f.].

As an immediate consequence of spontaneous chiral-symmetry breaking, the $\text{SO}\!\left(6
\right)$-symmetric wave-function renormalization $Z_{k}$ splits into two separate 
contributions associated with the $\boldsymbol{\pi}$ fields and the $\sigma$,
\begin{align}
	&Z^{\boldsymbol{\pi}}_{k} = Z_{k} + 2\sigma^{2}Z_{2,k} - 2\sigma^{2}p^{2}\!\left(C_{6,k}+
	C_{8,k}\right) \! , \label{eq:zp} \\
	&Z^{\sigma}_{k} = Z_{k} + 2\sigma^{2}\!\left(C_{2,k}+Z_{2,k}\right) \nonumber \\
	&\qquad\; -2\sigma^{2}p^{2}\!\left(C_{6,k}+C_{7,k}+C_{8,k}\right) \! , \label{eq:zs}
\end{align}
where $p$ refers to the external momentum from the functional derivatives w.r.t.\
the bosonic fields; cf.\ the Appendix. Using the above expressions and the 
fermionic wave-function renormalization factor, we define the renormalized 
(physical) fields and model parameters as
\begin{IEEEeqnarray}{rCl}
	\tilde{\boldsymbol{\pi}} & = &
	\sqrt{Z^{\boldsymbol{\pi}}_{k}}\boldsymbol{\pi}, \quad
	\tilde{\sigma} = \sqrt{Z^{\boldsymbol{\pi}}_{k}}\sigma, \quad
	\tilde{\psi} = \sqrt{Z^{\psi}_{k}}\psi, \nonumber\\
	\tilde{h}_{\text{ESB}} & = & \frac{h_{\text{ESB}}}{\sqrt{Z^{\boldsymbol{\pi}}_{k}}}, \quad
	\tilde{y}_{k} = \frac{y_{k}}{\sqrt{Z^{\boldsymbol{\pi}}_{k}}Z^{\psi}_{k}}, 
	\label{eq:renormquants} \\
	\tilde{C}_{i,k} & = & \frac{C_{i,k}}{(Z^{\boldsymbol{\pi}}_{k})^{2}}, \quad
	i = 1, \ldots, 8, \quad 
	\tilde{Z}_{2,k} = \frac{Z_{2,k}}{(Z^{\boldsymbol{\pi}}_{k})^{2}}. \nonumber
\end{IEEEeqnarray}
The renormalized quantities will lead to a well-normal\-ized kinetic term of the 
effective action in the upcoming Section. For the sake of simplicity, Eqs.\ (\ref{eq:zp}) 
and (\ref{eq:zs}) are evaluated at $p = 0$ during the flow. The coupling $C_{1,k}$ 
in Eq.\ (\ref{eq:renormquants}) corresponds to the momentum-independent quartic 
interaction in the effective potential $U_{k}$, see also the Appendix.

\subsection{Nonlinear effective pseudo-Nambu-Goldstone action}
\label{sec:NGaction}

The low-energy limit of the $\text{SO}\!\left(6\right)$-invariant QMDM is a nonlinear effective model 
over the vacuum manifold $\text{SU}\!\left(4\right)\!/\text{Sp}\!\left(4\right)\cong\text{SO}\!\left(
6\right)\!/\text{SO}\!\left(5\right)$. It is generated by eliminating the ``heavy'' 
dynamics of all non-pNGB fields from the renormalized effective average action. To this 
end, along the lines of Ref.\ \cite{Divotgey:2019xea}, we start with the fermionic sector 
of the QMDM and drop all operators depending on the quark fields $\psi$, as their respective 
contributions to the low-energy couplings are already captured by the FRG integration and
only tree-level amplitudes are necessary to cover quantum effects (based on $\Gamma$). 
Afterwards, we are left with a purely bosonic model that has the same structure as
the usual $\text{SO}\!\left(6\right)$ Linear Sigma Model.

Now, in order to eliminate the massive isoscalar component $\sigma$ of the vector $\varphi$ 
in Eq.\ (\ref{eq:phi}), we make use of the diffeomorphism between the vacuum manifold and 
the five-sphere, $\text{SO}\!\left(6\right)\!/\text{SO}\!\left(5\right)\cong S^{5}$, and 
rotate the renormalized bosonic field configuration into a frame that yields a clear 
separation between its light and massive modes,
\begin{equation}
	\begin{aligned}
	\tilde{\varphi} \equiv &\ \sqrt{Z^{\boldsymbol{\pi}}_{k}}\varphi = 
	\Sigma\bigl(\tilde{\zeta}\;\!\bigr)\tilde{\phi}, \\[0.1cm] 
	\tilde{\phi} = &\ \bigl(\boldsymbol{0}, \tilde{\theta}\;\!\bigr).
	\end{aligned}
	\label{eq:so6rot}
\end{equation}
The field $\tilde{\theta} = \sqrt{\tilde{\varphi}\cdot\tilde{\varphi}}$ describes the radial 
excitation of Eq.\ (\ref{eq:phi}). In contrast, the pNGB d.o.f.\ constitute a set of 
coordinates on $S^{5}$ and are written as $\tilde{\zeta}^{a}$, $a = 1,\ldots,5$. From the 
technical viewpoint, the transformation $\Sigma\bigl(\tilde{\zeta}\;\!\bigr)$ associates 
each point on $\text{SO}\!\left(6\right)\!/\text{SO}\!\left(5\right)$, i.e., each (left) 
coset, with an element of $\text{SO}\!\left(6\right)$ and is thus interpreted as a coset 
representative. Choosing stereographic projections as coordinates,
\begin{equation}
	\tilde{\zeta}^{a} = \frac{\tilde{\boldsymbol{\pi}}^{a}}{\tilde{\theta} + \tilde{\sigma}}, 
	\quad a = 1,\ldots,5,
	\label{eq:coords}
\end{equation}
its explicit form is given by
\begin{equation}
	\Sigma\bigl(\tilde{\zeta}\;\!\bigr) = 
	\begin{pmatrix}
		\delta^{a}_{\;\;b} - \frac{2\tilde{\zeta}^{a}\tilde{\zeta}_{b}}{1 + \tilde{\zeta}^{2}} &
		\frac{2\tilde{\zeta}^{a}}{1 + \tilde{\zeta}^{2}} \\[0.2cm]
		-\frac{2\tilde{\zeta}_{b}}{1 + \tilde{\zeta}^{2}} & \frac{1 - \tilde{\zeta}^{2}}{1 + \tilde{
		\zeta}^{2}}
	\end{pmatrix} \! .
	\label{eq:cosetrep}
\end{equation}

Apart from its geometric meaning, the coset representative can also be used to define 
further objects, such as the Maurer-Cartan form $\alpha_{\mu}$,
\begin{equation}
	\alpha_{\mu}\bigl(\tilde{\zeta}\;\!\bigr) = \Sigma\bigl(
	\tilde{\zeta}\;\!\bigr)^{-1}\partial_{\mu}
	\Sigma\bigl(\tilde{\zeta}\;\!\bigr) .
	\label{eq:mcform}
\end{equation}
It takes its values in the Lie algebra $\mathfrak{so}\!\left(6\right)$ and 
therefore admits an expansion of the form
\begin{equation}
	\alpha_{\mu}\bigl(\tilde{\zeta}\;\!\bigr) = ie^{\;\;a}_{\alpha}\bigl(\tilde{\zeta}\;\!\bigr)
	\partial_{\mu}\tilde{\zeta}^{\alpha}x_{a} + i\omega^{\;\;i}_{\alpha}\bigl(\tilde{\zeta}
	\;\!\bigr)\partial_{\mu}\tilde{\zeta}^{\alpha}s_{i},
	\label{eq:mcexp}
\end{equation}
where $\alpha = 1,\ldots,5$ represents a ``curved'' coset index and $x_{a}$, $a = 1,\ldots,5$, 
as well as $s_{i}$, $i = 1,\ldots,10$, denote the broken and unbroken generators of $\text{SO}\!
\left(6\right)$, respectively. The 15 generators $J^{a}_{\;b}$ (skew-symmetric matrices) 
in the fundamental representation fulfill the relation
\begin{equation}
	\left(J^{a}_{\;\;b}\right)^{m}_{\;\;n} = -i\left(\delta^{am}\delta_{bn}
	- \delta^{a}_{\;\;n}\delta^{m}_{\;\;b}\right), 
\end{equation}
with the indices $a,b,m,n = 1, \ldots, 6$, $a < b$, and $x_{a} = J^{a}_{\;\;6}$. These 
considerations are generalizations of the generators and transformation properties of 
the geometric objects quoted in Ref.\ \cite{Divotgey:2019xea}.
\begin{figure*}[t!]
	\centering
		\includegraphics{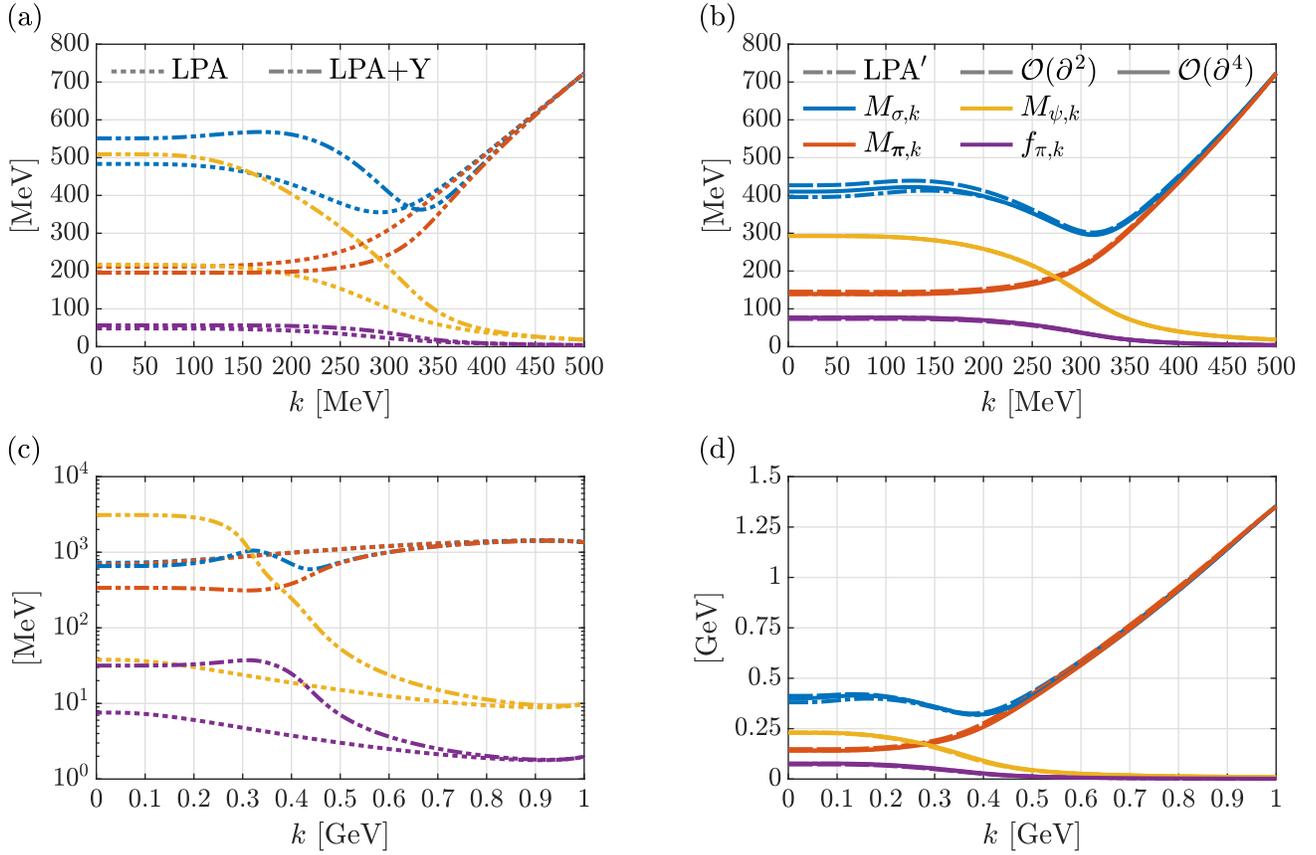}
	\caption{Scale-dependent renormalized masses and the pion decay constant in 
	various truncations of the $\text{SO}\!\left(6\right)$ QMDM. Two different UV-cutoff 
	scenarios are shown: $\Lambda_{\mathrm{UV}} = 500\ \mathrm{MeV}$ [subfigures (a) 
	and (b)] and $\Lambda_{\mathrm{UV}} = 1\ \mathrm{GeV}$ [subfigures
	(c) and (d)]; the legends of subfigures (a) and (b) apply to all panels.}
	\label{fig:masses}
\end{figure*}
\begin{table*}[t!]
	\caption{\label{tab:IRpars}IR parameters. All parameters are 
	evaluated at $k = k_{\mathrm{IR}} = 0.5\ \text{MeV}$.}
	\begin{ruledtabular}
		\begin{tabular}{lrrrrrrrrrr}
		& \multicolumn{5}{c}{$\boldsymbol{\Lambda_{\mathrm{UV}} = 500}\ \mathbf{MeV}$} 
		& \multicolumn{5}{c}{$\boldsymbol{\Lambda_{\mathrm{UV}} = 1}\ \mathbf{GeV}$} \\
		\cline{2-6} \cline{7-11}\\[-0.2cm]
		\textbf{Parameter} & \multicolumn{1}{c}{$\mathcal{O}\!\left(\partial^{4}\right)$} & 
		\multicolumn{1}{c}{$\mathcal{O}\!\left(\partial^{2}\right)$} & 
		\multicolumn{1}{c}{$\text{LPA}^{\prime}$} & \multicolumn{1}{c}{LPA+Y} & 
		\multicolumn{1}{c}{LPA} & 
		\multicolumn{1}{c}{$\mathcal{O}\!\left(\partial^{4}\right)$} & 
		\multicolumn{1}{c}{$\mathcal{O}\!\left(\partial^{2}\right)$} & 
		\multicolumn{1}{c}{$\text{LPA}^{\prime}$} & \multicolumn{1}{c}{LPA+Y} & 
		\multicolumn{1}{c}{LPA} \\[0.1cm]
		\colrule\\[-0.3cm]
		$M_{\boldsymbol{\pi}}\ [\mathrm{MeV}]$ & 139.0 & 145.5 & 139.1 & 195.7 & 211.5 & 
		140.8 & 148.6 & 141.9 & 338.8 & 693.0 \\[0.1cm]
		$M_{\sigma}\ [\mathrm{MeV}]$ & 410.1 & 426.9 & 395.6 & 551.0 & 483.3 & 
		400.9 & 412.7 & 380.5 & 656.2 & 725.5 \\[0.1cm]
		$M_{\psi}\ [\mathrm{MeV}]$ & 292.6 & 293.0 & 292.9 & 509.0 & 216.9 & 
		230.4 & 228.5 & 230.7 & 3106.9 & 37.8 \\[0.1cm]
		$f_{\pi}\ [\mathrm{MeV}]$ & 76.9 & 73.3 & 76.8 & 56.3 & 48.2 & 
		76.9 & 72.2 & 76.2 & 31.7 & 7.6 \\[0.1cm]
		$\tilde{y}$ & 3.81 & 4.00 & 3.81 & 9.04 & 4.50 & 
		2.99 & 3.16 & 3.03 & 98.08 & 5.00 \\[0.1cm]
		$Z^{\boldsymbol{\pi}}$ & 2.10 & 1.93 & 2.10 & 1.00 & 1.00 & 
		5.68 & 5.19 & 5.62 & 1.00 & 1.00 \\[0.1cm]
		$Z^{\sigma}$ & 1.78 & 1.62 & 1.92 & 1.00 & 1.00 & 
		5.70 & 5.19 & 6.29 & 1.00 & 1.00 \\[0.1cm]
		$Z^{\psi}$ & 1.17 & 1.18 & 1.17 & 1.00 & 1.00 & 
		1.24 & 1.24 & 1.24 & 1.00 & 1.00 \\[0.1cm]
		$\tilde{\alpha}_{1}/\!\left(10f_{\pi}^{2}\right)$ & 0.16 & 0.20 & 0.16 & 0.38 & 1.39 & 
		0.17 & 0.24 & 0.18 & 6.04 & $7.39 \times 10^{2}$ \\[0.1cm]
		$\tilde{C}_{1}$ & $-2.61$ & $-3.07$ & $-2.61$ & $-9.46$ & $-9.92$ & 
		$-2.99$ & $-3.61$ & $-3.08$ & $-40.93$ & $-90.65$ \\[0.1cm]
		$\tilde{\alpha}_{3}f_{\pi}^{2}$ & $-0.56$ & $-0.61$ & $-0.58$ & 26.09 & $-3.58$ & 
		2.78 & 2.80 & 2.72 & 80.85 & 0.17 \\[0.1cm]
		$\tilde{\alpha}_{4}f_{\pi}^{4}/10$ & $-1.04$ & $-1.19$ & $-1.03$ & $-16.22$ & $-0.83$ & 
		$-1.11$ & $-1.00$ & $-1.07$ & $6.05$ & $0.01$ \\[0.1cm]
		$\tilde{\alpha}_{5}f_{\pi}^{6}/10^{2}$ & 0.79 & 0.86 & 0.78 & 4.11 & 0.41 & 
		0.69 & 0.55 & 0.65 & $-2.16$ & $-8.58\times 10^{-5}$ \\[0.1cm]
		$\tilde{\alpha}_{6}f_{\pi}^{8}/10^{3}$ & $-0.73$ & $-0.75$ & $-0.71$ & 4.81 & $-0.22$ & 
		$-0.60$ & $-0.41$ & $-0.55$ & $-1.11$ & $1.50\times 10^{-6}$ \\
		\end{tabular}
	\end{ruledtabular}
\end{table*}

In the following, we will mainly be interested in the coefficients proportional to 
the broken generators [the coefficients $\omega^{\;\;i}_{\alpha}\bigl(\tilde{\zeta}\bigr)$
are not of further interest here], i.e.,
\begin{equation}
	e^{\;\;a}_{\alpha}\bigl(\tilde{\zeta}\;\!\bigr) = 
	\frac{2\delta_{\alpha}^{\;\;a}}{1 + \tilde{\zeta}^{2}}.
	\label{eq:vielbein}
\end{equation}
This identity derives from Eqs.\ (\ref{eq:cosetrep}) and (\ref{eq:mcexp}). The
coefficients $e^{\;\;a}_{\alpha}\bigl(\tilde{\zeta}\bigr)$ define a vielbein on 
$\text{SO}\!\left(6\right)\!/\text{SO}\!\left(5\right)$ and give rise to the metric
\begin{equation}
	g_{\alpha\beta}\bigl(\tilde{\zeta}\;\!\bigr) \equiv 
	\delta_{ab}e^{\;\;a}_{\alpha}\bigl(\tilde{\zeta}
	\;\!\bigr)e^{\;\;b}_{\beta}\bigl(\tilde{\zeta}\;\!\bigr) = 
	\frac{4\delta_{\alpha\beta}}{\bigl(1 + \tilde{\zeta}^{2}\bigr)^{2}}.
	\label{eq:metric}
\end{equation}

Finally, inserting Eq.\ (\ref{eq:so6rot}) into the remaining Linear Sigma Model and 
restricting the bosonic dynamics to the vacuum manifold, i.e., setting the radial 
excitation to the pion decay constant $f_{\pi}$,
\begin{equation}
	\tilde{\theta} = f_{\pi},
	\label{eq:restr}
\end{equation}
the renormalized effective pNGB action takes the following form:
\begin{align}
	\Gamma_{\! k} &= \int_{x}\, \biggl[\,\frac{f_{\pi}^{2}}{2}g_{\alpha\beta}\bigl(\nabla_{\mu}
	\tilde{\zeta}^{\alpha}\bigr)\nabla_{\mu}\tilde{\zeta}^{\beta} \nonumber\\[-0.1cm]
	&\qquad\;\;\;\,\,- \bigl(\tilde{C}_{6,k}+\tilde{C}_{8,k}\bigr)f_{\pi}^{4}g_{\alpha\beta}
	\bigl(\nabla_{\mu}\nabla_{\mu}\tilde{\zeta}^{\alpha}\bigr)\nabla_{\nu}\nabla_{\nu}\tilde{
	\zeta}^{\beta} \nonumber \\
	&\qquad\;\;\;\,\,- \bigl(\tilde{C}_{3,k} - \tilde{C}_{5,k}+\tilde{C}_{6,k}+\tilde{C}_{7,k}+
	\tilde{C}_{8,k}\bigr)f^{4}_{\pi} \nonumber \\
	&\qquad\qquad\;\;\;\,\,\,\times g_{\alpha\beta}g_{\gamma\delta}\bigl(\nabla_{\mu}\tilde{
	\zeta}^{\alpha}\bigr)\bigl(\nabla_{\mu}\tilde{\zeta}^{\beta}\bigr)\bigl(\nabla_{\nu}\tilde{
	\zeta}^{\gamma}\bigr)\nabla_{\nu}\tilde{\zeta}^{\delta} \nonumber \\ 
	&\qquad\;\;\;\,\,- \tilde{C}_{4,k}f^{4}_{\pi}g_{\alpha\beta}g_{\gamma\delta}\bigl(\nabla_{
	\mu}\tilde{\zeta}^{\alpha}\bigr)\bigl(\nabla_{\nu}\tilde{\zeta}^{\beta}\bigr)\bigl(\nabla_{
	\mu}\tilde{\zeta}^{\gamma}\bigr)\nabla_{\nu}\tilde{\zeta}^{\delta} \nonumber \\
	&\qquad\;\;\;\,\,-\tilde{h}_{\text{ESB}}f_{\pi}\frac{1-\tilde{\zeta}^{2}}{1+\tilde{\zeta}^{2}
	}\,\biggr], 
	\label{eq:nlsm}
\end{align}
where the action of the covariant derivative $\nabla_{\mu}$ onto the pNGB fields as 
well as their space-time derivatives is defined in the usual way, see Ref.\
\cite{Divotgey:2019xea}. The above equation resembles the most general ansatz 
for the derivative expansion of the $\text{SO}\!\left(6\right)$ Nonlinear 
Sigma Model up to (and including) $\mathcal{O}\!\left(\partial^{4}\right)$ 
\cite{Percacci:2009fh, Flore:2012ma}. However, as a consequence of explicit 
chiral-symmetry breaking, Eq.\ (\ref{eq:nlsm}) also features a potential for 
the pNGB fields.

Similar to Ref.\ \cite{Divotgey:2019xea}, the low-energy limit of the QMDM is 
obtained by expanding Eq.\ (\ref{eq:nlsm}) up to fourth order in the rescaled 
pNGB fields $\smash{\tilde{\Pi}^{a} = 2f_{\pi}\tilde{\zeta}^{a}}$, $a = 1,\ldots, 
5$; one arrives at
\begin{align}
	\Gamma_{\! k} &= \int_{x}\, \biggl\{ \, \frac{1}{2}\Bigl(\partial_{\mu}\tilde{\Pi}_{a}\Bigr)
	\partial_{\mu}\tilde{\Pi}^{a} + \frac{1}{2}\tilde{\mathcal{M}}^{2}_{\Pi,k}\tilde{\Pi}_{a}
	\tilde{\Pi}^{a} \nonumber \\[-0.1cm]
	&\qquad\;\;\;\, -\tilde{\mathcal{C}}_{1,k}\Bigl(\tilde{\Pi}_{a}\tilde{\Pi}^{a}\Bigr)^{2} + 
	\tilde{\mathcal{Z}}_{2,k}\tilde{\Pi}_{a}\tilde{\Pi}^{a}\Bigl(\partial_{\mu}\tilde{\Pi}_{b}
	\Bigr)\partial_{\mu}\tilde{\Pi}^{b} \nonumber \\
	&\qquad\;\;\;\, -\tilde{\mathcal{C}}_{3,k}\Bigl[\Bigl(\partial_{\mu}\tilde{\Pi}_{a}\Bigr)
	\partial_{\mu}\tilde{\Pi}^{a}\Bigr]^{2} \nonumber \\
	&\qquad\;\;\;\, -\tilde{\mathcal{C}}_{4,k}\Bigl[\Bigl(\partial_{\mu}\tilde{\Pi}_{a}\Bigr)
	\partial_{\nu}\tilde{\Pi}^{a}\Bigr]^{2} \nonumber \\
	&\qquad\;\;\;\, -\tilde{\mathcal{C}}_{5,k}\tilde{\Pi}_{a}\Bigl(\partial_{\mu}\partial_{\mu}
	\tilde{\Pi}^{a}\Bigr)\Bigl(\partial_{\nu}\tilde{\Pi}_{b}\Bigr)\partial_{\nu}\tilde{\Pi}^{b}
	\nonumber \\
	&\qquad\;\;\;\, -\tilde{\mathcal{C}}_{6,k}\tilde{\Pi}_{a}\tilde{\Pi}^{a}\Bigl(\partial_{\mu}
	\partial_{\nu}\tilde{\Pi}_{b}\Bigr)\partial_{\mu}\partial_{\nu}\tilde{\Pi}^{b}\nonumber \\[-0.1cm]
	&\qquad\;\;\;\, -\tilde{\mathcal{C}}_{8,k}\tilde{\Pi}_{a}\tilde{\Pi}^{a}\Bigl(\partial_{\mu}
	\partial_{\mu}\tilde{\Pi}_{b}\Bigr)\partial_{\nu}\partial_{\nu}\tilde{\Pi}^{b}\biggr\} ,
	\label{eq:lowelim}
\end{align}
where we defined the (squared) pNGB mass as
\begin{equation}
	\tilde{\mathcal{M}}^{2}_{\Pi,k} = \frac{\tilde{h}_{\text{ESB}}}{f_{\pi}}.
	\label{eq:pngbmass}
\end{equation}
The low-energy couplings of the QMDM in Eq.\ (\ref{eq:lowelim}) are functions of the 
linear model parameters and the pion decay constant $f_{\pi}$,
\begin{equation}
\begin{aligned}
	\tilde{\mathcal{C}}_{1,k} = &\ \frac{\tilde{\mathcal{M}}^{2}_{\Pi,k}}{8f^{2}_{\pi}}, \\
	\tilde{\mathcal{Z}}_{2,k} = &\ -\frac{1}{4f^{2}_{\pi}}, \\
	\tilde{\mathcal{C}}_{3,k} = &\ \tilde{C}_{3,k} - \tilde{C}_{5,k} + \tilde{C}_{7,k} 
	+ 2\bigl(\tilde{C}_{6,k} + \tilde{C}_{8,k}\bigr), \\
	\tilde{\mathcal{C}}_{4,k} = &\ \tilde{C}_{4,k}, \\
	\tilde{\mathcal{C}}_{5,k} = &\ 2\bigl(\tilde{C}_{6,k}+\tilde{C}_{8,k}\bigr), \\
	\tilde{\mathcal{C}}_{6,k} = &\ -\tilde{C}_{6,k}-\tilde{C}_{8,k}, \\
	\tilde{\mathcal{C}}_{8,k} = &\ \frac{1}{2}\bigl(\tilde{C}_{6,k}+\tilde{C}_{8,k}\bigr).
	\label{eq:lecs}
\end{aligned}
\end{equation}
We observe that the geometric constraint of fixing the dynamics of the linear
model to the five-sphere eliminates the couplings $\tilde{\mathcal{C}}_{2,k}$ and 
$\tilde{\mathcal{C}}_{7,k}$. Additionally, the couplings $\tilde{\mathcal{C}}_{5,k}$,
$\tilde{\mathcal{C}}_{6,k}$, and $\tilde{\mathcal{C}}_{8,k}$ exhibit a linear dependence 
w.r.t.\ the sum $\tilde{C}_{6,k} + \tilde{C}_{8,k}$.

\section{Results}
\label{sec:results}

In the present qualitative study, we shed light on the FRG scaling behavior 
of higher-derivative couplings of the $\text{SO}\!\left(6\right)$-invariant 
QMDM. The UV values of the scale-dependent variables in Eq.\ (\ref{eq:qmdmtrunc}) are 
tuned such that physical values for the meson and (constituent) quark masses are produced 
in the IR limit, if possible [following ``usual'' QCD \cite{Tanabashi:2018oca}]: 
$M_{\boldsymbol{\pi}} \simeq 139\ \mathrm{MeV}$, $M_{\sigma} \simeq 400 - 500\ \mathrm{MeV}$, 
and $M_{\psi} \simeq 300 - 330\ \mathrm{MeV}$ (about one third of the proton mass). However, the 
desired value for the pion decay constant, $f_{\pi} \simeq 76\ \mathrm{MeV}$, which is related 
to the scale-dependent minimum of the effective potential $U_{k}$, is $\sqrt{N_{c}}$-scaled 
as compared to QCD ($f_{\pi} \simeq 93\ \mathrm{MeV}$). The IR values of the higher-derivative 
interactions, which are initialized at zero in the UV, are understood as model predictions.

Besides the explicit investigation of different truncations of the QMDM [LPA,
LPA+Y (LPA and the flow of the Yukawa coupling $y_{k}$), $\text{LPA}^{\prime}$, 
couplings at $\mathcal{O}\!\left(\partial^{2}\right)$, and couplings at 
$\mathcal{O}\!\left(\partial^{4}\right)$], we choose two different UV-cutoff scenarios, 
namely, $\Lambda_{\mathrm{UV}} = 500\ \mathrm{MeV}$ and $\Lambda_{\mathrm{UV}} = 1\ \mathrm{GeV}$. 
These two scenarios give insights into possible tuning procedures and provide a first
(rather simple and qualitative) estimate of UV-cutoff dependences of the higher-derivative 
interactions and, hence, the low-energy couplings of the QMDM. Both chosen scales are 
frequently used in FRG studies of the model \cite{Strodthoff:2011tz, Khan:2015puu}.
\begin{figure*}[t!]
	\centering
		\includegraphics{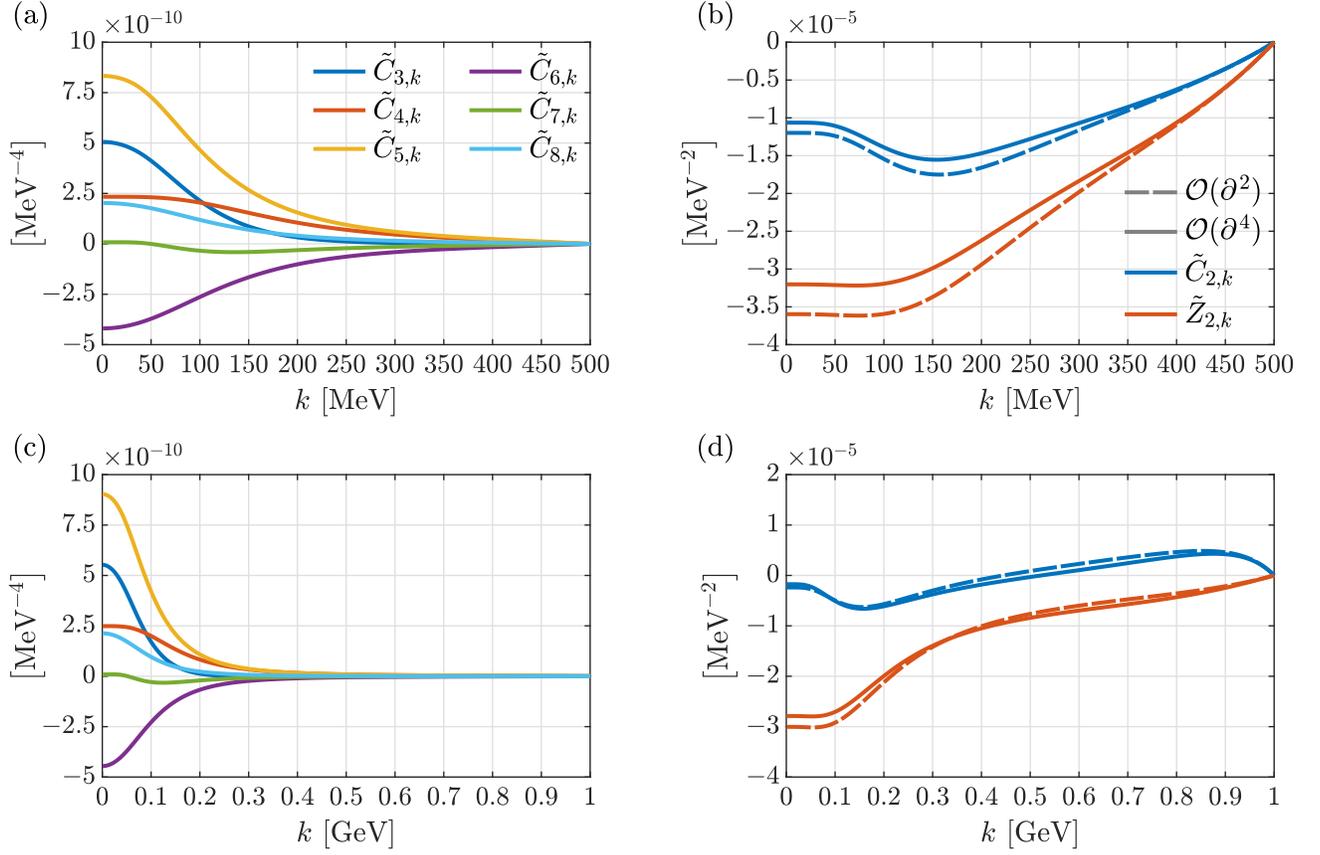}
	\caption{Flow of higher-derivative couplings of the $\text{SO}\!\left(6\right)$ 
	QMDM (linear symmetry realization). The legends in subfigures (a) and (b) apply 
	to all panels.}
	\label{fig:cp24}
\end{figure*}

The FRG flow is stopped at $k \equiv k_{\mathrm{IR}} = 0.5\ \mathrm{MeV}$ to reduce 
numerical costs; at $k = k_{\mathrm{IR}}$, there is no longer any scale dependence. 
Further details about the FRG flow equations and their numerical solutions are 
deferred to the Appendix.
\begin{figure*}[t!]
	\centering
		\includegraphics{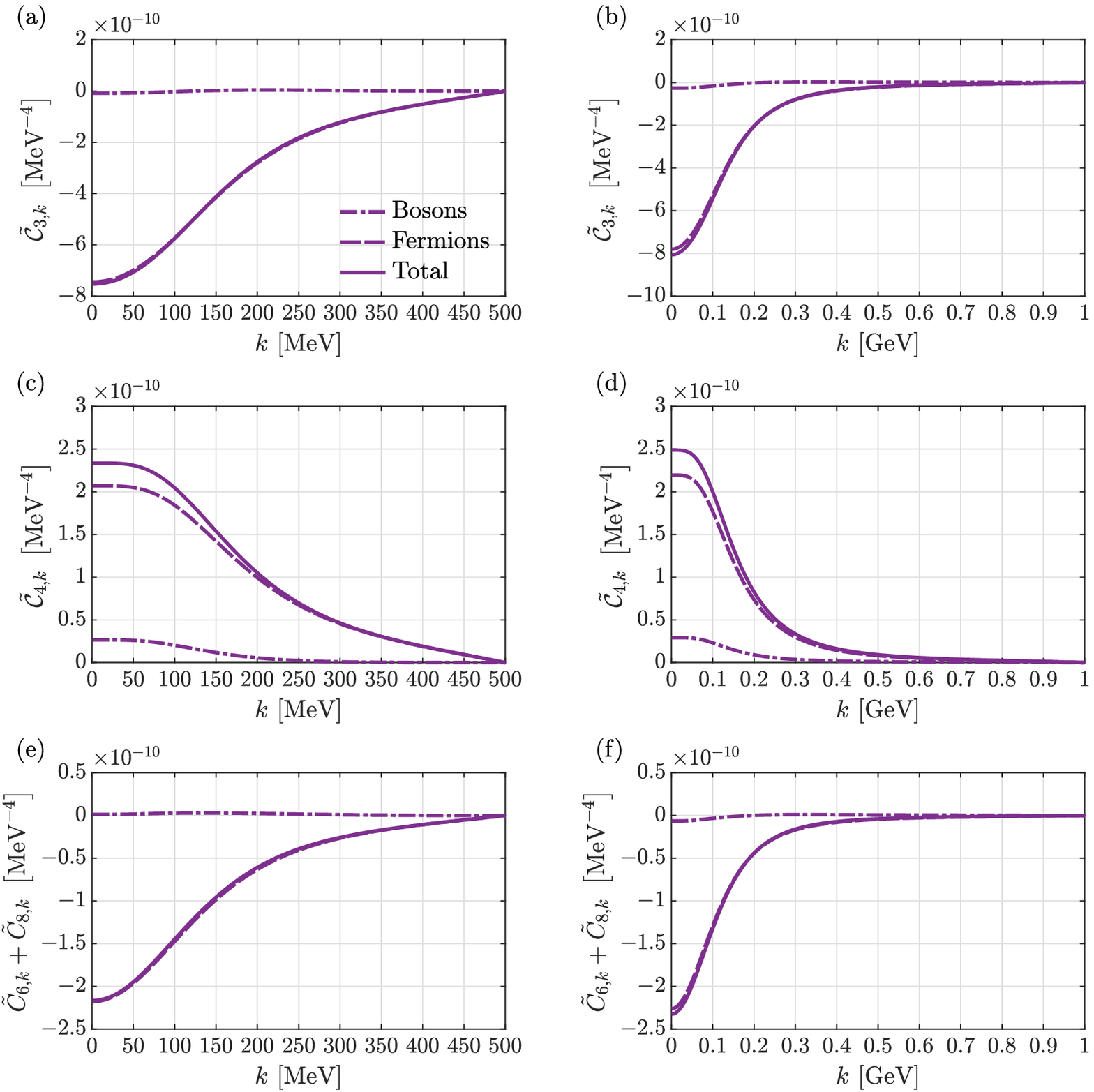}
	\caption{FRG scale evolution of the low-energy couplings of the 
	$\text{SO}\!\left(6\right)$-symmetric QMDM. The total flows (solid lines)
	are divided into contributions from bosonic and fermionic quantum fluctuations
	(dash-dotted and dashed lines, respectively). The legend in subfigure (a)
	is valid for all panels. Since the couplings $\tilde{\mathcal{C}}_{5,k}$,
	$\tilde{\mathcal{C}}_{6,k}$, and $\tilde{\mathcal{C}}_{8,k}$ are linearly
	dependent according to Eq.\ (\ref{eq:lecs}), we only show the sum 
	$\tilde{C}_{6,k} + \tilde{C}_{8,k}$ in subfigures (e) and (f).}
	\label{fig:cnonlinear}
\end{figure*}

\subsection{Linear model:\\Flow of the higher-derivative couplings}
\label{sec:linearmodel}

The scale-dependent squared meson and quark masses are derived from the effective
potential and the Yukawa interaction as
\begin{equation}
	m_{\boldsymbol{\pi},k}^{2} = 2U_{k}', \quad
	m_{\sigma,k}^{2} = 2U_{k}' + 4\rho U_{k}'', \quad
	m_{\psi,k}^{2} = y_{k}^{2}\rho .
\end{equation}
These expressions are evaluated at the scale-dependent minimum $\rho_{0,k}$ 
of the term $U_{k} - h_{\mathrm{ESB}}\sqrt{\rho}$. The corresponding renormalized
masses are given by
\begin{equation}
	M_{\boldsymbol{\pi},k}^{2} = \frac{m_{\boldsymbol{\pi},k}^{2}}
	{Z_{k}^{\boldsymbol{\pi}}}, \quad
	M_{\sigma,k}^{2} = \frac{m_{\sigma,k}^{2}}{Z_{k}^{\sigma}}, \quad 
	M_{\psi,k}^{2} = \frac{m_{\psi,k}^{2}}{(Z_{k}^{\psi})^{2}}.
\end{equation}
For the two different scenarios, we were able to find the following values for
the (renormalized) masses and the pion decay constants:
\begin{itemize}
\item $M_{\boldsymbol{\pi},k_{\mathrm{IR}}} = 139.0\ \mathrm{MeV}$, $M_{\sigma,k_{\mathrm{IR}}} 
= 410.1\ \mathrm{MeV}$, and $M_{\psi,k_{\mathrm{IR}}} = 292.6\ \mathrm{MeV}$, as well as
$f_{\pi} = 76.9\ \mathrm{MeV}$ ($\Lambda_{\mathrm{UV}} = 500\ \mathrm{MeV}$);
\item $M_{\boldsymbol{\pi},k_{\mathrm{IR}}} = 140.8\ \mathrm{MeV}$, $M_{\sigma,k_{\mathrm{IR}}} 
= 400.9\ \mathrm{MeV}$, and $M_{\psi,k_{\mathrm{IR}}} = 230.4\ \mathrm{MeV}$, as well as
$f_{\pi} = 76.9\ \mathrm{MeV}$ ($\Lambda_{\mathrm{UV}} = 1\ \mathrm{GeV}$).
\end{itemize}
These findings are collected in Table \ref{tab:IRpars} in the column 
``$\mathcal{O}\!\left(\partial^{4}\right)$'' (for both chosen UV cutoffs), which stands
for the ``full'' flow, i.e., we take into account all higher-derivative couplings 
introduced in Eq.\ (\ref{eq:qmdmtrunc}). We remark that the generated masses and 
decay constants are in the ``physical'' ranges mentioned above, except for
the quark masses, which appear to be slightly to light. The fact that it was harder
to find reasonable values for the larger cutoff of 1 GeV already points 
out limitations of the model and (possibly) exceeded validity ranges.
\begin{table*}[t!]
	\caption{\label{tab:LECslinear}Higher-derivative couplings (linear 
	model), meson and quark masses, and pion decay constants. The scale 
	of initialization of the higher-derivative interactions, $\Lambda_{C}$, 
	is given below the (general) cutoff scale $\Lambda_{\mathrm{UV}}$, 
	$\Lambda_{C} = \text{factor} \times \Lambda_{\mathrm{UV}}$. All
	couplings, masses, and decay constants are evaluated at $k = k_{\mathrm{IR}} 
	= 0.5\ \text{MeV}$.}
	\begin{ruledtabular}
		\begin{tabular}{lrrrrrrrrrrrrrrr}
		& \multicolumn{5}{c}{$\boldsymbol{\Lambda_{\mathrm{UV}} = 500}\ \mathbf{MeV}$} 
		& \multicolumn{10}{c}{$\boldsymbol{\Lambda_{\mathrm{UV}} = 1}\ \mathbf{GeV}$} \\
		\cline{2-6} \cline{7-16}\\[-0.3cm]
		\textbf{Parameter} & 1.0 & 0.8 & 0.6 & 0.4 & 0.2 & 
		1.0 & 0.9 & 0.8 & 0.7 & 0.6 & 0.5 & 0.4 & 0.3 & 0.2 & 0.1 \\
		\colrule\\[-0.3cm]
		$M_{\boldsymbol{\pi}}\ [\mathrm{MeV}]$ & 
		139.0 & 139.2 & 139.2 & 139.1 & 139.1 & 
		140.8 & 140.9 & 141.1 & 141.3 & 141.5 & 141.7 & 141.8 & 142.0 & 142.0 & 141.9 \\[0.1cm]
		$M_{\sigma}\ [\mathrm{MeV}]$ & 
		410.1 & 409.8 & 407.4 & 401.3 & 395.9 & 
		400.9 & 400.6 & 400.5 & 400.2 & 399.7 & 398.8 & 396.9 & 393.2 & 386.4 & 380.8 \\[0.1cm]
		$M_{\psi}\ [\mathrm{MeV}]$ & 
		292.6 & 292.8 & 292.8 & 292.9 & 292.9 & 
		230.4 & 230.5 & 230.6 & 230.6 & 230.6 & 230.6 & 230.6 & 230.7 & 230.7 & 230.7 \\[0.1cm]
		$f_{\pi}\ [\mathrm{MeV}]$ & 
		76.9 & 76.8 & 76.7 & 76.8 & 76.8 & 
		76.9 & 76.9 & 76.7 & 76.6 & 76.5 & 76.3 & 76.2 & 76.1 & 76.1 & 76.2 \\[0.1cm]
		$\tilde{C}_{2}\ [1/f_{\pi}^{2}]\times 10$ & 
		$-0.63$ & $-0.48$ & $-0.18$ & 0.26 & 0.24 & 
		$-0.10$ & $-0.11$ & $-0.11$ & $-0.11$ & $-0.10$ & $-0.08$ & $-0.02$ & 0.11 & 0.33 & 0.23 \\[0.1cm]
		$\tilde{Z}_{2}\ [1/f_{\pi}^{2}]\times 10$ & 
		$-1.89$ & $-1.68$ & $-1.26$ & $-0.56$ & $-0.01$ & 
		$-1.65$ & $-1.65$ & $-1.64$ & $-1.63$ & $-1.60$ & $-1.56$ & $-1.45$ & 
		$-1.22$ & $-0.71$ & $-0.06$ \\[0.1cm]
		$\tilde{C}_{3}\ [1/f_{\pi}^{4}]\times 10^{2}$ & 
		1.77 & 1.80 & 1.86 & 1.87 & 1.15 & 
		1.94 & 1.93 & 1.93 & 1.92 & 1.92 & 1.92 & 1.94 & 1.98 & 2.04 & 1.43 \\[0.1cm]
		$\tilde{C}_{4}\ [1/f_{\pi}^{4}]\times 10^{2}$ & 
		0.82 & 0.78 & 0.68 & 0.44 & 0.08 &
		0.87 & 0.87 & 0.87 & 0.86 & 0.85 & 0.84 & 0.82 & 0.76 & 0.58 & 0.15 \\[0.1cm]
		$\tilde{C}_{5}\ [1/f_{\pi}^{4}]\times 10^{2}$ & 
		2.91 & 2.90 & 2.86 & 2.62 & 1.42 &
		3.16 & 3.15 & 3.14 & 3.12 & 3.11 & 3.10 & 3.09 & 3.09 & 2.96 & 1.79 \\[0.1cm]
		$\tilde{C}_{6}\ [1/f_{\pi}^{4}]\times 10^{2}$ & 
		$-1.47$ & $-1.44$ & $-1.38$ & $-1.17$ & $-0.55$ &
		$-1.56$ & $-1.56$ & $-1.55$ & $-1.55$ & $-1.54$ & $-1.54$ & $-1.53$ & 
		$-1.49$ & $-1.35$ & $-0.75$ \\[0.1cm]
		$\tilde{C}_{7}\ [1/f_{\pi}^{4}]\times 10^{2}$ & 
		0.03 & 0.04 & 0.07 & 0.16 & 0.18 & 
		0.03 & 0.03 & 0.02 & 0.02 & 0.02 & 0.02 & 0.02 & 0.04 & 0.12 & 0.16 \\[0.1cm]
		$\tilde{C}_{8}\ [1/f_{\pi}^{4}]\times 10^{2}$ & 
		0.71 & 0.70 & 0.69 & 0.61 & 0.30 &
		0.75 & 0.75 & 0.74 & 0.74 & 0.74 & 0.74 & 0.74 & 0.73 & 0.69 & 0.41 \\
		\end{tabular}
	\end{ruledtabular}
\end{table*}

\begin{table*}[t!]
	\caption{\label{tab:LECsnonlinear}Low-energy couplings (nonlinear model). 
	The scale of initialization of the higher-derivative 
	interactions, $\Lambda_{C}$, is given below the (general) cutoff scale 
	$\Lambda_{\mathrm{UV}}$, $\Lambda_{C} = \text{factor} \times \Lambda_{\mathrm{UV}}$. 
	All couplings are evaluated at $k = k_{\mathrm{IR}} = 0.5\ \text{MeV}$.
	The renormalized pion mass $\tilde{\mathcal{M}}_{\Pi,k}$ is given in Table 
	\ref{tab:LECslinear}, $\tilde{\mathcal{M}}_{\Pi,k} \equiv M_{\boldsymbol{\pi},k}$.}
	\begin{ruledtabular}
		\begin{tabular}{lddddddddddddddd}
		& \multicolumn{5}{c}{$\boldsymbol{\Lambda_{\mathrm{UV}} = 500}\ \mathbf{MeV}$} 
		& \multicolumn{10}{c}{$\boldsymbol{\Lambda_{\mathrm{UV}} = 1}\ \mathbf{GeV}$} \\
		\cline{2-6} \cline{7-16}\\[-0.3cm]
		\textbf{Coupling} & 1.0 & 0.8 & 0.6 & 0.4 & 0.2 & 
		1.0 & 0.9 & 0.8 & 0.7 & 0.6 & 0.5 & 0.4 & 0.3 & 0.2 & 0.1 \\
		\colrule\\[-0.3cm]
		$\tilde{\mathcal{C}}_{1}\times 10$ & 
		4.09 & 4.10 & 4.11 & 4.10 & 4.09 & 
		4.18 & 4.20 & 4.22 & 4.25 & 4.28 & 4.31 & 4.33 & 4.35 & 4.35 & 4.34 \\[0.1cm]
		$\tilde{\mathcal{Z}}_{2}\ [1/f_{\pi}^{2}]\times 10$ & 
		-2.50 & -2.50 & -2.50 & -2.50 & -2.50 & 
		-2.50 & -2.50 & -2.50 & -2.50 & -2.50 & -2.50 & -2.50 & -2.50 & -2.50 & -2.50 \\[0.1cm]
		$\tilde{\mathcal{C}}_{3}\ [1/f_{\pi}^{4}]\times 10^{2}$ & 
		-2.63 & -2.55 & -2.32 & -1.71 & -0.59 & 
		-2.83 & -2.82 & -2.81 & -2.80 & -2.78 & -2.76 & -2.71 & -2.58 & -2.13 & -0.88 \\[0.1cm]
		$\tilde{\mathcal{C}}_{4}\ [1/f_{\pi}^{4}]\times 10^{2}$ & 
		0.82 & 0.78 & 0.68 & 0.44 & 0.08 &
		0.87 & 0.87 & 0.87 & 0.86 & 0.85 & 0.84 & 0.82 & 0.76 & 0.58 & 0.15 \\[0.1cm]
		$\tilde{\mathcal{C}}_{5}\ [1/f_{\pi}^{4}]\times 10^{2}$ & 
		-1.51 & -1.48 & -1.39 & -1.12 & -0.49 &
		-1.63 & -1.63 & -1.62 & -1.61 & -1.60 & -1.59 & -1.58 & -1.52 & -1.33 & -0.69 \\[0.1cm]
		$\tilde{\mathcal{C}}_{6}\ [1/f_{\pi}^{4}]\times 10^{2}$ & 
		0.76 & 0.74 & 0.69 & 0.56 & 0.25 &
		0.82 & 0.81 & 0.81 & 0.81 & 0.80 & 0.80 & 0.79 & 0.76 & 0.67 & 0.34 \\[0.1cm]
		$\tilde{\mathcal{C}}_{8}\ [1/f_{\pi}^{4}]\times 10^{2}$ & 
		-0.38 & -0.37 & -0.35 & -0.28 & -0.12 &
		-0.41 & -0.41 & -0.41 & -0.40 & -0.40 & -0.40 & -0.39 & -0.38 & -0.33 & -0.17 \\
		\end{tabular}
	\end{ruledtabular}
\end{table*}

Concerning the different truncation schemes, we plot in Fig.\ \ref{fig:masses}
the entire evolution of the scale-dependent masses and the pion decay constant 
for all approximations (the starting values in the UV are kept constant). It reveals 
that the largest corrections (deviations) come from the inclusion (neglection) of 
the FRG flow of the wave-function renormalization, cf.\ Fig.\ \ref{fig:masses}(a) 
in comparison to Fig.\ \ref{fig:masses}(b). Also, the flow of the Yukawa coupling 
$y_{k}$ has a relatively large influence on the outcome in the IR, which confirms 
the study in Ref.\ \cite{Khan:2015puu}; thus the inclusion of wave-function 
renormalizations is of major importance for a consistent truncation scheme at 
this specific momentum-independent level.

These observations are even more drastic in the cutoff scenario of 1 GeV, where 
in the ``lower'' truncations LPA and LPA+Y the results are partly of a different 
order of magnitude, cf. Fig.\ \ref{fig:masses}(c) and Fig.\ \ref{fig:masses}(d). In 
particular, chiral symmetry cannot even be considered as (substantially) broken in 
the LPA truncation; the quantities $M_{\boldsymbol{\pi},k_{\mathrm{IR}}}$ and 
$M_{\sigma,k_{\mathrm{IR}}}$ only differ by less than 40 MeV and the decay constant 
stays below 10 MeV, see the last column in Table \ref{tab:IRpars}. In the ``higher''
truncations [$\text{LPA}^{\prime}$, $\mathcal{O}\!\left(\partial^{2}\right)$, and
$\mathcal{O}\!\left(\partial^{4}\right)$], the energy-momentum scale of chiral-symmetry 
breaking is shifted from around $300 - 400\ \mathrm{MeV}$ to larger values of 
$400 - 500\ \mathrm{MeV}$ [Fig.\ \ref{fig:masses}(b) and Fig.\ \ref{fig:masses}(d)],
when increasing the UV cutoff from 500 MeV to 1 GeV. The breaking is signaled by the growing 
mass split between the pNGB fields and the $\sigma$ occurring in this region. One further 
finds that the observables within the $\text{LPA}^{\prime}$ truncation are in some instances 
good approximations to the final results of $\mathcal{O}\!\left(\partial^{4}\right)$,
e.g.\ for the pion mass listed in the first row of Table \ref{tab:IRpars} or the pion 
decay constant in the fourth row. This is consistent with Ref.\ \cite{Eser:2018jqo} and 
could be explained by cancellation effects in the wave-function renormalization 
factors (\ref{eq:zp}) and (\ref{eq:zs}) beyond the $\text{LPA}^{\prime}$.

The explicit FRG flows of the wave-function renormalization factors and the Yukawa 
coupling are found in the Appendix. Moreover, Table \ref{tab:IRpars} contains 
their IR-limit values as well as the renormalized Taylor coefficients 
$\tilde{\alpha}_{n,k}$, $n = 1,\ldots , 6$, which were used to numerically solve 
the flow equation for the effective potential $U_{k}$; see once more the Appendix
for details. The coupling $\tilde{\alpha}_{2,k}$ is replaced by its identification
(\ref{eq:c1}) with the (momentum-independent) quartic interaction. At this point, 
we register that these coefficients reflect again the large differences in the 
effective potential w.r.t.\ the various investigated truncations.
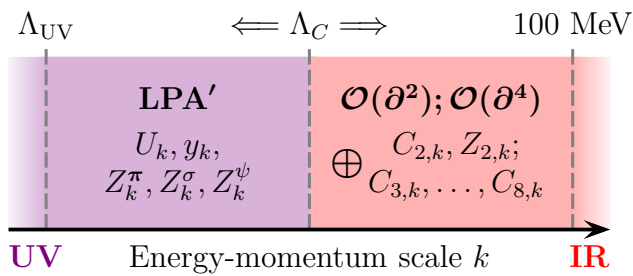
\begin{figure}[t!]
	\centering
	\begin{pspicture}[showgrid=false](-4.3,-0.6)(4.3,3.4)
		\psframe[fillstyle=solid,fillcolor=violet!30,linestyle=none](-3.6,0)(0.1,2.3)
		\psframe[fillstyle=solid,fillcolor=red!30,linestyle=none](0,0)(3.6,2.3)
		\psframe[fillstyle=gradient,gradbegin=red!30,gradend=red!5,gradmidpoint=1.0,
		gradangle=90,linestyle=none](3.5,0)(4,2.3)
		\psframe[fillstyle=gradient,gradbegin=violet!30,gradend=violet!5,gradmidpoint=1.0,
		gradangle=-90,linestyle=none](-4,0)(-3.5,2.3)
		\psline[linewidth=0.04,linecolor=gray,linestyle=dashed,dash=6pt 2pt](-3.5,0)(-3.5,2.4)
		\psline[linewidth=0.04,linecolor=gray,linestyle=dashed,dash=6pt 2pt](0,0)(0,2.4)
		\psline[linewidth=0.04,linecolor=gray,linestyle=dashed,dash=6pt 2pt](3.5,0)(3.5,2.4)
		\psline[linewidth=0.05,arrowsize=3pt 3]{->}(-4,0)(4,0)
		\rput[B](-3.5,2.6){\large $\Lambda_{\mathrm{UV}}$}
		\rput[B](0,2.6){\large $\Longleftarrow \Lambda_{C}\,\! \Longrightarrow$}
		\rput[B](3.5,2.6){\large 100 MeV}
		\rput[B](0,-0.5){\large Energy-momentum scale $k$}
		\rput[Bl](-4,-0.5){\large\color{violet}{\textbf{UV}}}
		\rput[Br](4,-0.5){\large\color{red}{\textbf{IR}}}
		\rput[B](-1.75,1.6){\large $\boldsymbol{\mathrm{LPA}^{\prime}}$}
		\rput[B](-1.75,1){\large $U_{k},y_{k},$}
		\rput[B](-1.75,0.5){\large $Z_{k}^{\boldsymbol{\pi}},Z_{k}^{\sigma},Z_{k}^{\psi}$}
		\rput[B](1.75,1.6){\large $\boldsymbol{\mathcal{O}\!\left(\partial^{2}\right)\! ;%
		\mathcal{O}\!\left(\partial^{4}\right)}$}
		\rput[B](1.95,1){\large $C_{2,k},Z_{2,k};$}
		\rput[B](1.95,0.5){\large $C_{3,k},\ldots , C_{8,k}$}
		\rput[B](0.5,0.75){\Large $\boldsymbol{\oplus}$}
	\end{pspicture}
	\caption{Study of cutoff dependences. The cutoff $\Lambda_{C}$ for the
	higher-derivative couplings $C_{2,k}$, $Z_{2,k}$, and $C_{i,k}$, $i = 3,\ldots ,8$, 
	varies between 100 MeV and the overall UV cutoff $\Lambda_{\mathrm{UV}}$ (in 
	steps of 100 MeV each). These enter the dynamics at $\Lambda_{C}$ in addition 
	to $(\oplus)$ the $\mathrm{LPA}^{\prime}$ truncation (i.e., the flow of the effective 
	potential $U_{k}$, the Yukawa coupling $y_{k}$, as well as the wave-function 
	renormalization factors $Z_{k}^{\boldsymbol{\pi}}$, $Z_{k}^{\sigma}$, and 
	$Z_{k}^{\psi}$); the $\text{LPA}^{\prime}$ flow is initialized at 
	$\Lambda_{\mathrm{UV}}$.}
	\label{fig:cutoff}
\end{figure}

Figure \ref{fig:cp24} demonstrates how the higher-derivative couplings are (exclusively)
generated from quantum fluctuations, starting from zero at $\Lambda_{\mathrm{UV}} = 
500\ \mathrm{MeV}$ [Fig.\ \ref{fig:cp24}(a) and Fig.\ \ref{fig:cp24}(b)] and 
$\Lambda_{\mathrm{UV}} = 1\ \mathrm{GeV}$ [Fig.\ \ref{fig:cp24}(c) and Fig.\ \ref{fig:cp24}(d)].
Here, we decided to present the $\mathcal{O}\!\left(\partial^{4}\right)$-couplings first
as they are more relevant for the low-energy limit in the nonlinear realization, cf.\ 
Eq.\ (\ref{eq:lecs}); the $\mathcal{O}\!\left(\partial^{2}\right)$-couplings do not 
enter the limit explicitly. Comparing the upper to the lower panels, we notice that
the evolution of the renormalized higher-derivative couplings becomes ``flatter'' in the
high-energy regime for the scenario with a larger UV cutoff. Such a behavior was already 
anticipated in the study of the $\text{SO}\!\left(4\right)$-symmetric QMM \cite{Divotgey:2019xea} 
and it underlines their character as couplings parametrizing low-energy operators, e.g.\ 
the couplings $\tilde{C}_{i,k}$, $i = 3, \ldots , 8$, in Fig.\ \ref{fig:cp24}(c) become 
solely relevant below $k \simeq 500\ \mathrm{MeV}$. The relatively steep slope of the curves
close to $\Lambda_{\mathrm{UV}}$ in Fig.\ \ref{fig:cp24}(b) and Fig.\ \ref{fig:cp24}(d) 
indicates, however, that the couplings $\tilde{C}_{2,k}$ and $\tilde{Z}_{2,k}$ should 
still be present in the UV. Finally, the effect of the FRG flow of the couplings of 
$\mathcal{O}\!\left(\partial^{4}\right)$ onto these $\mathcal{O}\!\left(\partial^{2}\right)$-couplings 
shrinks from Fig.\ \ref{fig:cp24}(b) to Fig.\ \ref{fig:cp24}(d) (see the dashed and 
solid lines). The precise IR values of the higher-derivative couplings are discussed 
in Sec.\ \ref{sec:cutoff}, together with a qualitative estimate of their cutoff dependences.
\begin{figure*}[t!]
	\centering
		\includegraphics{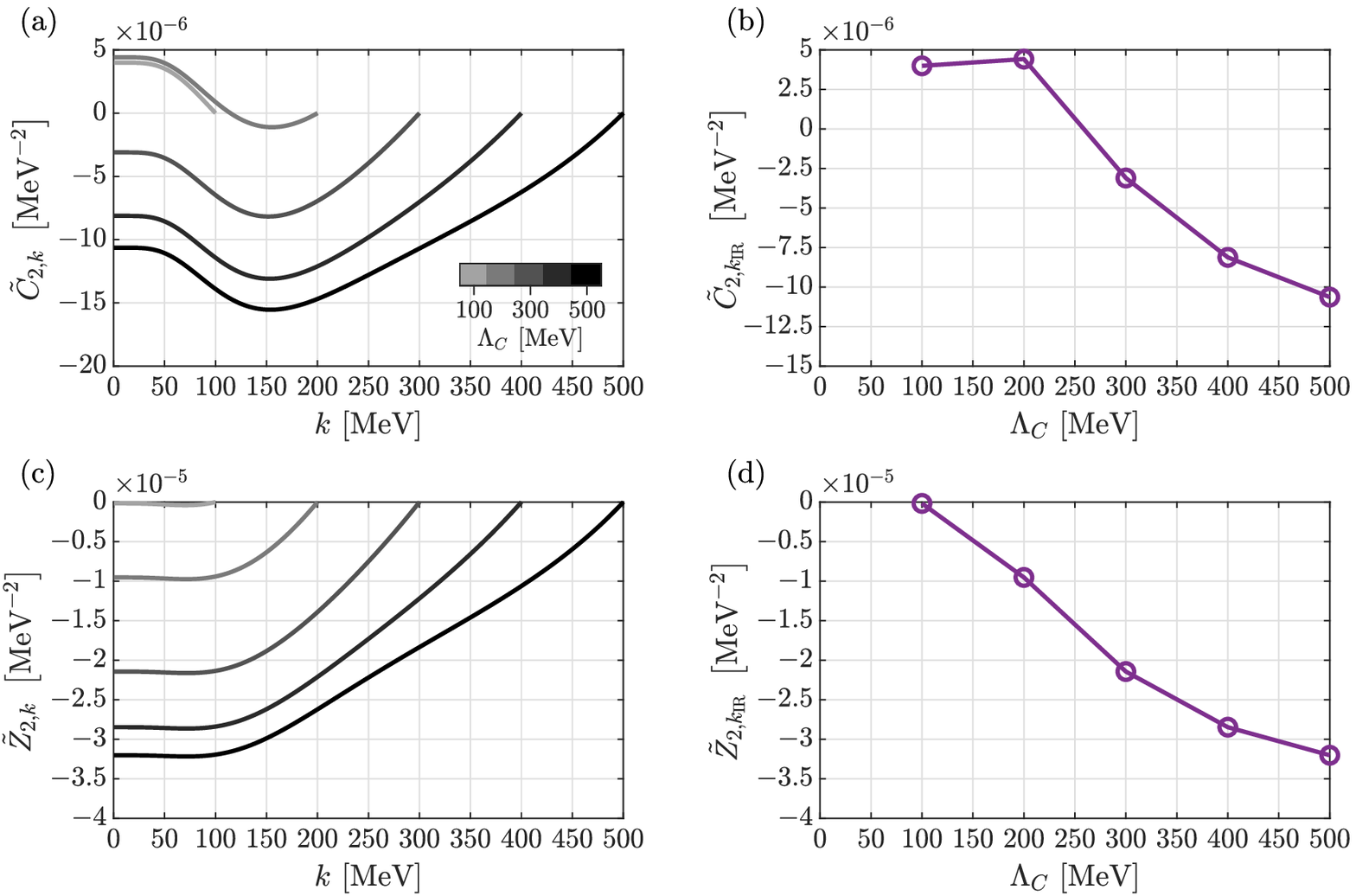}
	\caption{Scale evolution of the higher-derivative couplings $\tilde{C}_{2,k}$
	and $\tilde{Z}_{2,k}$ as a function of the initialization scale $\Lambda_{C}$.
	The flow was tuned at $\Lambda_{\mathrm{UV}} = 500\ \mathrm{MeV}$. The IR-limit 
	values are shown in subfigures (b) and (d), respectively. The legend in 
	subfigure (a) is valid for the panels (a) and (c).}
	\label{fig:cp2_UV_500}
\end{figure*}

\subsection{Nonlinear model:\\Transition to the low-energy couplings}
\label{sec:nonlinearmodel}

A physically meaningful transition scale to the nonlinear model is reached 
upon a decoupling of all heavy d.o.f.\ (i.e., heavier than the pNGBs) 
from the FRG flow. Nevertheless, Eq.\ (\ref{eq:lecs}) allows us to map the linear 
model with its higher-derivative interactions on all energy-momentum scales onto 
the nonlinear counterpart, which grants full access to the entire scaling behavior 
of the low-energy couplings.
\begin{figure*}[t!]
	\centering
		\includegraphics{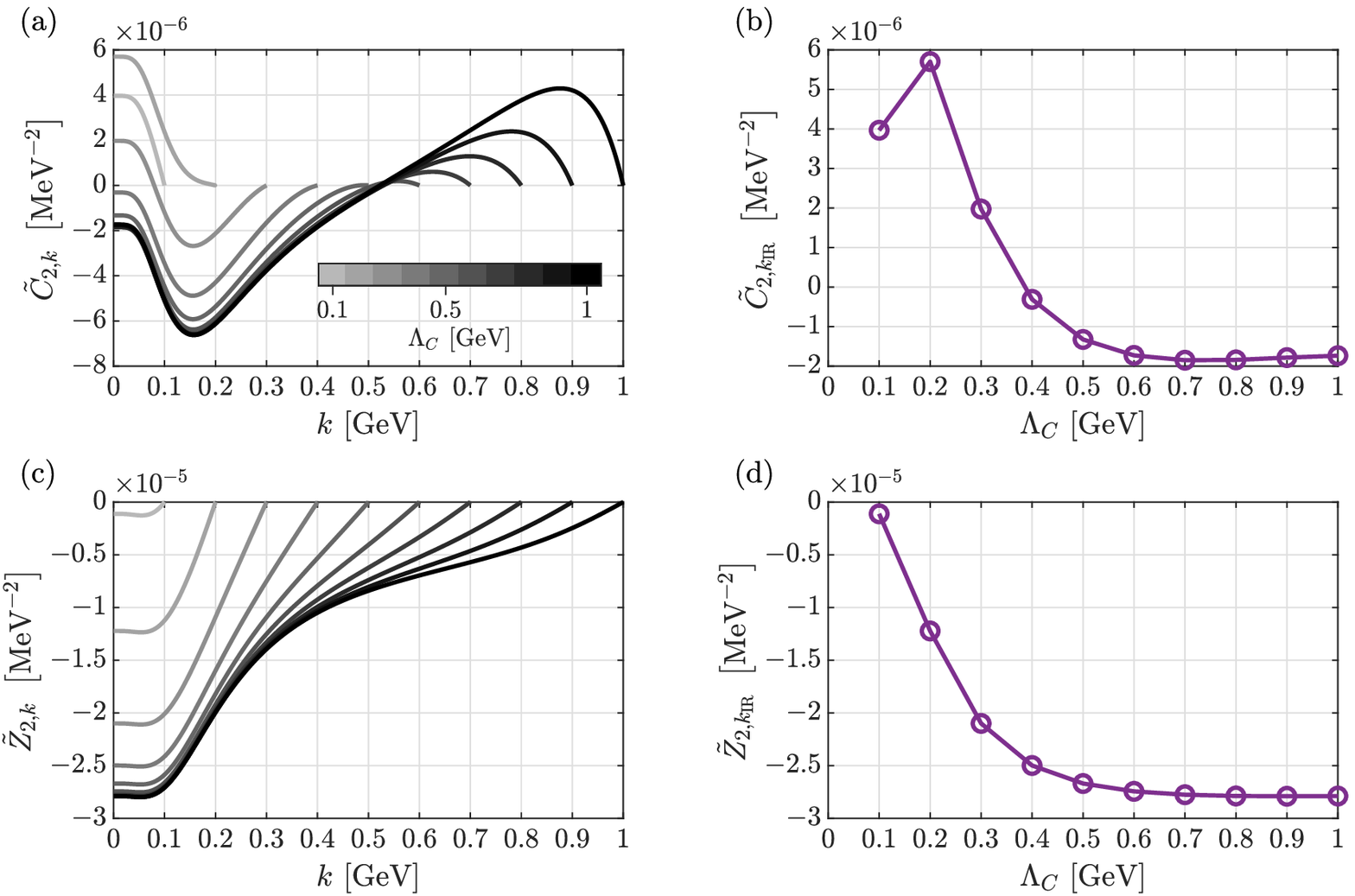}
	\caption{Scale evolution of the higher-derivative couplings $\tilde{C}_{2,k}$
	and $\tilde{Z}_{2,k}$ as a function of the initialization scale $\Lambda_{C}$.
	The flow was tuned at $\Lambda_{\mathrm{UV}} = 1\ \mathrm{GeV}$. The IR-limit 
	values are shown in subfigures (b) and (d), respectively. The legend in 
	subfigure (a) is valid for the panels (a) and (c).}
	\label{fig:cp2_UV_1000}
\end{figure*}

Figure \ref{fig:cnonlinear} illustrates the mapping onto the nonlinear pNGB
action (\ref{eq:lowelim}), carried out on all scales $k$ within the respective
UV-cutoff scenario. The flow of the low-energy couplings $\tilde{\mathcal{C}}_{3,k}$,
$\tilde{\mathcal{C}}_{4,k}$, and the combination $\tilde{C}_{6,k} + \tilde{C}_{8,k}$
[representing the linearly dependent low-energy couplings $\tilde{\mathcal{C}}_{5,k}$,
$\tilde{\mathcal{C}}_{6,k}$, and $\tilde{\mathcal{C}}_{8,k}$, cf.\ Eq.\ (\ref{eq:lecs})]
is thereby split into bosonic and fermionic loop contributions. The results exhibit a
clear dominance of the fermionic quantum fluctuations as compared to the bosonic ones, 
astonishingly, even for relatively low energies below the scale of chiral-symmetry breaking 
($300 - 400\ \mathrm{MeV}$ and $400 - 500\ \mathrm{MeV}$ for $\Lambda_{\mathrm{UV}} = 
500\ \mathrm{MeV}$ and $\Lambda_{\mathrm{UV}} = 1\ \mathrm{GeV}$, respectively). 
This illustration nicely demonstrates the naturally emerging fluctuation hierarchy 
of quark loops at larger momenta and the pNGB fluctuations at low energy-momentum 
scales within the FRG. From the upcoming pNGB fluctuations and the onset of 
``flattening'' in the fermionic curve in the rough region of $50 - 150\ \mathrm{MeV}$ (for 
both cutoff scenarios) we infer an appropriate transition scale from the linear QMDM to 
its nonlinear low-energy limit of the same order, verifying the scales found in
Refs.\ \cite{Divotgey:2019xea, Eser:2019pvd}.
\begin{figure*}[t!]
	\centering
		\includegraphics{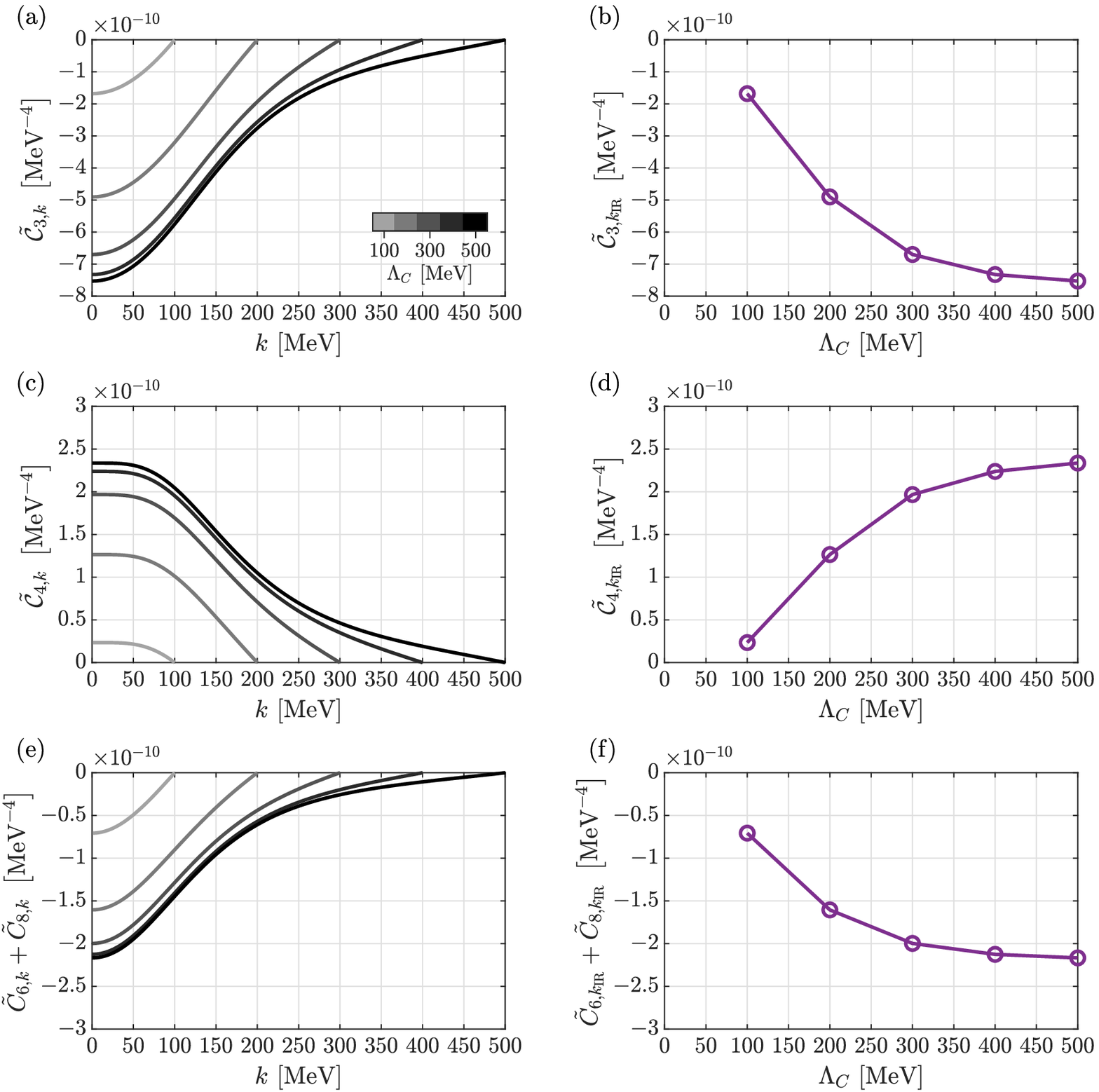}
	\caption{Scale evolution of the low-energy couplings $\tilde{\mathcal{C}}_{3,k}$
	and $\tilde{\mathcal{C}}_{4,k}$, as well as the sum $\tilde{C}_{6,k} + \tilde{C}_{8,k}$ 
	as a function of the initialization scale $\Lambda_{C}$. The flow was tuned at 
	$\Lambda_{\mathrm{UV}} = 500\ \mathrm{MeV}$. The IR-limit values are shown in 
	subfigures (b), (d), and (f), respectively. The legend in subfigure (a) applies
	to the panels (a), (c), and (e).}
	\label{fig:cp4_UV_500}
\end{figure*}

The geometrical constants $\tilde{\mathcal{C}}_{1,k}$ and $\tilde{\mathcal{Z}}_{2,k}$ 
are not shown in Fig.\ \ref{fig:cnonlinear}, since their scaling behavior is essentially 
dictated by the pion decay constant and the pion wave-function renormalization, see 
Eqs.\ (\ref{eq:pngbmass}) and (\ref{eq:lecs}). This means that they are fixed by the 
constant-radius condition (\ref{eq:restr}) on the five-sphere. The (scale-independent) 
explicit symmetry breaking parameter $h_{\mathrm{ESB}}$ in Eq.\ (\ref{eq:pngbmass}) is 
given in Table \ref{tab:UVpars} in the Appendix and we consistently find that 
$\tilde{\mathcal{M}}_{\Pi,k} = M_{\boldsymbol{\pi},k}$. As for the higher-derivative
interactions within the linearly realized QMDM in Sec.\ \ref{sec:linearmodel}, the 
IR values of the low-energy couplings in the nonlinear picture are discussed in 
Sec.\ \ref{sec:cutoff}.

The phenomenon of fermion dominance in these calculations is rather robust and 
was already observed in the $\text{SO}\!\left(4\right)$-symmetric QMM \cite{Divotgey:2019xea}. 
However, in this work and the present study within the QMDM, it comes up the question 
about the sensibility of the computed low-energy couplings in the IR limit w.r.t.\ changes 
in the UV cutoff.

\subsection{Cutoff dependences}
\label{sec:cutoff}

We qualitatively elucidate the UV-cutoff dependences of the linear higher-derivative 
and the nonlinear low-energy couplings with a rather simple setup, see Fig.\ 
\ref{fig:cutoff}. Besides the two UV cutoffs (and tuning scenarios) introduced 
above (for a first glimpse into truncation and cutoff effects), we define the 
intermediate scale $\Lambda_{C}$ at which we add the higher-derivative couplings 
to the $\text{LPA}^{\prime}$ flow. The latter is initialized at $\Lambda_{\mathrm{UV}}$
as before. Thus the presented curves in Sec.\ \ref{sec:linearmodel} and Sec.\ 
\ref{sec:nonlinearmodel} equal the case of $\Lambda_{C} \equiv \Lambda_{\mathrm{UV}}$. 
In order to qualitatively ``measure'' the cutoff dependence, the scale $\Lambda_{C}$ is 
successively lowered (in separate FRG runs) in steps of 100 MeV (down to 100 MeV as the 
lower bound).
\begin{figure*}[t!]
	\centering
		\includegraphics{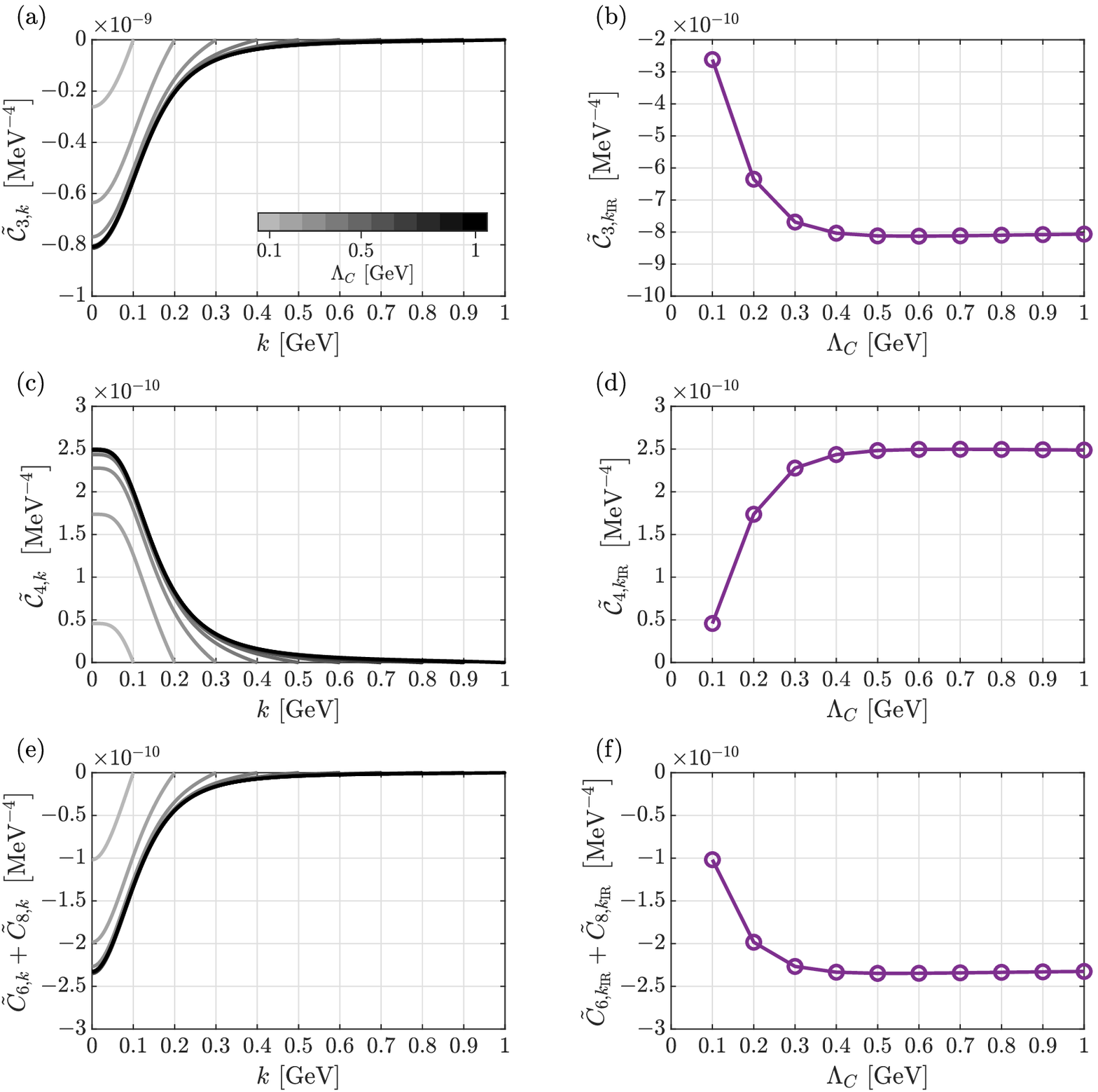}
	\caption{Scale evolution of the low-energy couplings $\tilde{\mathcal{C}}_{3,k}$
	and $\tilde{\mathcal{C}}_{4,k}$, as well as the sum $\tilde{C}_{6,k} + \tilde{C}_{8,k}$ 
	as a function of the initialization scale $\Lambda_{C}$. The flow was tuned at 
	$\Lambda_{\mathrm{UV}} = 1\ \mathrm{GeV}$. The IR-limit values are shown in 
	subfigures (b), (d), and (f), respectively. The legend in subfigure (a) applies
	to the panels (a), (c), and (e).}
	\label{fig:cp4_UV_1000}
\end{figure*}

This strategy is motivated by the fact that the $\text{LPA}^{\prime}$ flow is 
comparably ``constant'' w.r.t.\ the inclusion of the higher-derivative interactions, 
meaning that it is a reasonable approximation to the ``full'' truncation (as concluded 
above) and that distorting alterations in other observables are minimized. In fact, the 
deviation between those truncations in terms of the IR masses of the pNGBs and the quarks 
is less than $1\%$, cf.\ again Table \ref{tab:IRpars} [of course, one still has to consider 
the couplings of $\mathcal{O}\!\left(\partial^{2}\right)$ and $\mathcal{O}\!\left(\partial^{4}\right)$
to access the low-energy couplings of the model].

Tables \ref{tab:LECslinear} and \ref{tab:LECsnonlinear} show the results of this procedure
for the higher-derivative couplings (in the linear symmetry realization) and the low-energy
couplings (in the nonlinear model), respectively. The IR values of the flows in Fig.\ 
\ref{fig:cp24} and Fig.\ \ref{fig:cnonlinear} are listed in the second and seventh columns;
it is immediate that the order of magnitude is cutoff-independent, while, at the same time, 
the difference between the cutoff scenarios is partly large [i.e., for the couplings of
$\mathcal{O}\!\left(\partial^{2}\right)$ of the linear model]. However, the actually 
relevant couplings of $\smash{\mathcal{O}\!\left(\partial^{4}\right)}$ of the linear model as 
well as all couplings of the respective nonlinear model only differ by less than $10\%$.

Concerning the variation of $\Lambda_{C}$, we plot the different FRG flows of the 
higher-derivative couplings of $\mathcal{O}\!\left(\partial^{2}\right)$ in a gray-scaled
manner in Fig.\ \ref{fig:cp2_UV_500} ($\Lambda_{\mathrm{UV}} = 500\ \mathrm{MeV}$) and 
Fig.\ \ref{fig:cp2_UV_1000} ($\Lambda_{\mathrm{UV}} = 1\ \mathrm{GeV}$). The black lines
are identical to the respective curves in Fig.\ \ref{fig:cp24}. The FRG flows are further 
complemented by panels showing the IR values as a function of $\Lambda_{C}$ (in addition 
to the complete set of IR values quoted again in Tables \ref{tab:LECslinear} and 
\ref{tab:LECsnonlinear}). They reveal a persisting UV-cutoff dependence for larger 
$\Lambda_{\mathrm{C}}$ in the scenario of $\Lambda_{\mathrm{UV}} = 500\ \mathrm{MeV}$ 
[cf.\ Fig.\ \ref{fig:cp2_UV_500}(b) and Fig.\ \ref{fig:cp2_UV_500}(d)], but an almost 
``flat'' curve for $\Lambda_{\mathrm{UV}} = 1\ \mathrm{GeV}$ [cf.\ Fig.\ \ref{fig:cp2_UV_1000}(b) 
and Fig.\ \ref{fig:cp2_UV_1000}(d)]. 

The first four rows in Table \ref{tab:LECslinear} underline the ``robustness'' 
of the $\text{LPA}^{\prime}$ flow against the additional couplings of 
$\mathcal{O}\!\left(\partial^{2}\right)$ and $\mathcal{O}\!\left(\partial^{4}\right)$, 
underlying the variations of $\Lambda_{C}$. Solely the $\sigma$ mass changes by more 
than $1\%$, when decreasing or increasing $\Lambda_{C}$ (but still only by a maximum 
of around $5\%$); this justifies the setup of Fig.\ \ref{fig:cutoff} a posteriori.

Analogously, the couplings of $\mathcal{O}\!\left(\partial^{4}\right)$ are presented 
in Fig.\ \ref{fig:cp4_UV_500} and Fig.\ \ref{fig:cp4_UV_1000}, directly given in form 
of the low-energy couplings according to Eq.\ (\ref{eq:lecs}). For the IR-limit 
values as a function of the scale $\Lambda_{C}$, we find the same behavior
as for the $\mathcal{O}\!\left(\partial^{2}\right)$-couplings, i.e., the curves and 
values only ``converge'' for the case that $\Lambda_{\mathrm{UV}} = 1\ \mathrm{GeV}$. 
For initialization scales $\Lambda_{C} \gtrsim 800\ \mathrm{MeV}$, a reasonable
settlement of the IR values of both the $\mathcal{O}\!\left(\partial^{2}\right)$-couplings 
and those of $\smash{\mathcal{O}\!\left(\partial^{4}\right)}$ sets in. This finding matches
the scale proposed in Ref.\ \cite{Khan:2015puu} (in a similar thermodynamic consideration 
to ensure UV-cutoff independence of the results). Let us stress that this scale has to 
be further constrained by the proper determination of the validity range of the QMDM 
from the dynamical-hadronization technique along the lines of Refs.\ \cite{Gies:2001nw, 
Gies:2002hq, Pawlowski:2005xe, Floerchinger:2009uf, Braun:2014ata, Mitter:2014wpa, 
Khan:2015puu, Cyrol:2017ewj, Paris-Lopez:2018vjc, Alkofer:2018guy, Fu:2019hdw}.

\section{Summary and outlook}
\label{sec:summary}

The low-energy limit and qualitative features of the IR dynamics of 
the $\text{SO}\!\left(6\right)$-symmetric QMDM have been addressed within 
the FRG formalism. In order to determine the corresponding low-energy couplings, 
which have exclusively been generated from the FRG integration of quantum 
fluctuations, we considered a complete set of higher-derivative interactions
beyond traditional approximation schemes (such as LPA and $\text{LPA}^{\prime}$
truncations) and transformed the linear effective action into a nonlinear pNGB 
model; the pNGB fields themselves act as coordinates on the vacuum manifold 
$\text{SO}\!\left(6\right)\!/\text{SO}\!\left(5\right)$, which arises from 
the spontaneous breakdown of the chiral symmetry of $\text{QC}_{2}\text{D}$.

Consistent with recent studies in the $\text{SO}\!\left(4\right)$-symmetric 
QMM \cite{Eser:2018jqo, Divotgey:2019xea, Eser:2019pvd}, we found the following 
main results:
\begin{enumerate}
\item[(a)] The FRG flow of the low-energy couplings is heavily dominated 
by the quark fluctuations. The pNGB dynamics takes over below 
scales of $50 - 150\ \mathrm{MeV}$, which naturally marks a fluctuation-induced 
transition (and renormalization) scale between the linear QMDM to its nonlinearly 
realized low-energy limit (i.e., the computed effective pNGB action).

\item[(b)] The inclusion of higher-derivative interactions into the FRG flow 
demonstrates that LPA and LPA+Y truncations are rather insufficient, while its 
minimal extension ($\text{LPA}^{\prime}$) already takes major corrections into 
account.

\item[(c)] The UV-cutoff independence of the fluctuation-generated low-energy 
couplings, as it was investigated in a simple qualitative manner, impose cutoff 
scales of $\Lambda_{\mathrm{UV}} \gtrsim 800\ \mathrm{MeV}$. These scales are
further constrained in the broader context of determining the validity range 
of the QMDM [which is expected to be below such scales \cite{Braun:2014ata, 
Mitter:2014wpa, Khan:2015puu, Cyrol:2017ewj, Paris-Lopez:2018vjc, Alkofer:2018guy, 
Braun:2018svj, Fu:2019hdw} and highlights the complexity of finding appropriate 
cutoff scales]. Furthermore, it is questionable to what extent a ``retuning'' of
the UV parameters at every scale $\Lambda_{C}$ will change the results (albeit
the effect of higher-derivative couplings is small, cf.\ Tables \ref{tab:IRpars}
and \ref{tab:LECslinear}).
\end{enumerate}

This qualitative analysis is the basis for more involved investigations regarding
the validity range of the QMDM and a smooth transition from fundamental to low-energy 
operators. Especially, result (c) and the fact that the higher-derivative interactions
of $\mathcal{O}\!\left(\partial^{2}\right)$ experience large corrections close
to the UV-cutoff scale raise the question about possible distortions by the usage 
of ``corrupted'' d.o.f.\ in the high-energy domain. This necessitates a determination 
of the low-energy couplings from larger UV scales with the help of dynamical hadronization
\cite{Gies:2001nw, Gies:2002hq, Pawlowski:2005xe, Floerchinger:2009uf} as well as
a detailed discussion of cutoff (in)dependences in the generic sense of renormalization-group
consistency \cite{Braun:2018svj}. Nevertheless, the presented study follows the ultimate
goal of consistently generating low-energy couplings within the effective average action 
formalism, towards ``FRG-improved'' model calculations. It is understood as a 
complementary work to our recent studies within the $\text{SO}\!\left(4\right)$-invariant QMM.

Moreover, the effect of higher-derivative interactions should be studied at 
nonzero temperatures and densities. It would be interesting to learn how they change
the picture in the (expectedly very diverse) large-density regime, as an extension
(or addition) to Refs.\ \cite{Kogut:2000ek, Strodthoff:2011tz, Strodthoff:2013cua, 
Khan:2015puu, Leonhardt:2019fua}. Finally, the flow of the nonlinear effective 
pNGB action itself and the pion-mass dependence of the presented results will 
attract our attention in the near future.

\begin{acknowledgments}
The authors thank Dennis D.\ Dietrich, Adrian Koenigstein, Mario Mitter, 
Jan M.\ Pawlowski, Dirk H.\ Rischke, Bernd-Jochen Schaefer, and Lorenz
von Smekal for helpful discussions.
\end{acknowledgments}

\clearpage

\appendix*

\section{Flow equations}
\label{sec:floweqns}

We extract the flow equations of the scale-dependent quantities in the ansatz
(\ref{eq:qmdmtrunc}) from the Wetterich equation (\ref{eq:wetteq}) via functional 
differentiation w.r.t.\ the corresponding fields. The concrete expressions for
the flows of the effective potential $U_{k}$, the Yukawa coupling $y_{k}$, the 
wave-function renormalization factors $Z_{k}^{\boldsymbol{\pi}}$, $Z_{k}^{\sigma}$, 
and $\smash{Z_{k}^{\psi}}$, as well as the higher-derivative couplings $C_{2,k}$, $Z_{2,k}$, 
and $C_{i,k}$, $i = 3, \ldots , 8$, read in a diagrammatic language as follows:
\begin{widetext}
\vspace{-0.3cm}
\begin{IEEEeqnarray}{rCl}
	\partial_{k}U_{k} & = & \mathcal{V}^{-1} \partial_{k}\Gamma_{k}
	 = \mathcal{V}^{-1} \biggg( 
	\frac{1}{2} \! \! \vcenter{\hbox{

	}} \! \! \Biggg), \label{eq:c2} \\[1.0cm]
	\partial_{k}Z_{2,k} & = & \frac{1}{4} \mathcal{V}^{-1} 
	\left.\frac{\mathrm{d}}{\mathrm{d}p^2}\right|_{p^{2}\, =\, 0}
	\frac{\delta^{4}\partial_{k}\Gamma_{k}}{\delta\pi^{1}(p)
	\delta\pi^{2}(0)\delta\pi^{1}(-p)\delta\pi^{2}(0)} , \label{eq:z2} \\[1.0cm]
	\partial_{k}C_{3,k} & = & \frac{5}{576}\Bigg\lbrace 
	\frac{208}{5}\,\partial_{k}C_{5,k} - 112\, \partial_{k}C_{6,k}
	+ 32\, \partial_{k}C_{7,k} - \frac{224}{5}\, \partial_{k}C_{8,k}
	+ \mathcal{V}^{-1} \bigg(
	\frac{\partial}{\partial p_{1,\mu}}\frac{\partial}{\partial p_{2,\mu}}
	\frac{\partial}{\partial p_{3,\nu}}\frac{\partial}{\partial p_{1,\nu}} \nonumber\\
	& & \qquad\quad - \frac{7}{10}\frac{\partial}{\partial p_{1,\mu}}
	\frac{\partial}{\partial p_{2,\mu}}
	\frac{\partial}{\partial p_{3,\nu}}\frac{\partial}{\partial p_{2,\nu}}
	\bigg)\bigg|_{p_{1}\, =\, p_{2}\, =\, p_{3}\, =\, 0}
	\frac{\delta^{4}\partial_{k}\Gamma_{k}}{\delta\pi^{1}(p_{1})
	\delta\pi^{2}(p_{2})\delta\pi^{1}(p_{3})\delta\pi^{2}(-p_{1}-p_{2}-p_{3})}
	\Bigg\rbrace , \label{eq:c3} \\[1.0cm]
	\partial_{k}C_{4,k} & = & - \frac{1}{288}\Bigg\lbrace 
	- 16\, \partial_{k}C_{5,k} - 400\, \partial_{k}C_{6,k}
	+ 32\, \partial_{k}C_{7,k} - 160\, \partial_{k}C_{8,k}
	+ \mathcal{V}^{-1} \bigg(
	\frac{\partial}{\partial p_{1,\mu}}\frac{\partial}{\partial p_{2,\mu}}
	\frac{\partial}{\partial p_{3,\nu}}\frac{\partial}{\partial p_{1,\nu}} \nonumber\\
	& & \qquad\quad\ - \frac{5}{2}\frac{\partial}{\partial p_{1,\mu}}
	\frac{\partial}{\partial p_{2,\mu}}
	\frac{\partial}{\partial p_{3,\nu}}\frac{\partial}{\partial p_{2,\nu}}
	\bigg)\bigg|_{p_{1}\, =\, p_{2}\, =\, p_{3}\, =\, 0}
	\frac{\delta^{4}\partial_{k}\Gamma_{k}}{\delta\pi^{1}(p_{1})
	\delta\pi^{2}(p_{2})\delta\pi^{1}(p_{3})\delta\pi^{2}(-p_{1}-p_{2}-p_{3})}
	\Bigg\rbrace , \\[1.0cm]
	\partial_{k}C_{5,k} & = & \frac{1}{96} \mathcal{V}^{-1} 
	\bigg(\frac{\partial}{\partial p_{2,\mu}}\frac{\partial}{\partial p_{2,\mu}}
	\frac{\partial}{\partial p_{2,\nu}}\frac{\partial}{\partial p_{3,\nu}} \nonumber\\
	& & \qquad\qquad\quad - \frac{1}{2}\frac{\partial}{\partial p_{2,\mu}}
	\frac{\partial}{\partial p_{2,\mu}}
	\frac{\partial}{\partial p_{2,\nu}}\frac{\partial}{\partial p_{2,\nu}}
	\bigg)\bigg|_{p_{2}\, =\, p_{3}\, =\, 0}
	\frac{\delta^{4}\partial_{k}\Gamma_{k}}{\delta\pi^{1}(-p_{2}-p_{3})
	\delta\pi^{2}(p_{2})\delta\pi^{1}(p_{3})\delta\pi^{2}(0)}, \\[1.0cm]
	\partial_{k}C_{6,k} & = & - \frac{1}{96}\Bigg\lbrace 
	160\, \partial_{k}C_{5,k} + 64\, \partial_{k}C_{7,k} + \mathcal{V}^{-1}
	\frac{\partial}{\partial p_{2,\mu}}\frac{\partial}{\partial p_{4,\mu}}
	\frac{\partial}{\partial p_{2,\nu}}\frac{\partial}{\partial p_{4,\nu}}
	\bigg|_{p_{2}\, =\, p_{4}\, =\, 0}
	\frac{\delta^{4}\partial_{k}\Gamma_{k}}{\delta\pi^{1}(-p_{2}-p_{4})
	\delta\pi^{2}(p_{2})\delta\pi^{1}(0)\delta\pi^{2}(p_{4})} \nonumber\\
	& & \qquad\quad\, - \frac{1}{12} \mathcal{V}^{-1}
	\frac{\partial}{\partial p_{\mu}}
	\frac{\partial}{\partial p_{\mu}}
	\frac{\partial}{\partial p_{\nu}}\frac{\partial}{\partial p_{\nu}}
	\bigg|_{p\, =\, 0}
	\frac{\delta^{4}\partial_{k}\Gamma_{k}}{\delta\pi^{1}(-p)
	\delta\pi^{2}(0)\delta\pi^{1}(p)\delta\pi^{2}(0)}
	\Bigg\rbrace , \\[1.0cm]
	\partial_{k}C_{7,k} & = & - \frac{1}{384} \mathcal{V}^{-1}
	\frac{\partial}{\partial p_{\mu}}\frac{\partial}{\partial p_{\mu}}
	\frac{\partial}{\partial p_{\nu}}\frac{\partial}{\partial p_{\nu}}
	\bigg|_{p\, =\, 0}
	\frac{\delta^{4}\partial_{k}\Gamma_{k}}{\delta\pi^{1}(-p)
	\delta\pi^{2}(p)\delta\pi^{1}(0)\delta\pi^{2}(0)}, \\[1.0cm]
	\partial_{k}C_{8,k} & = & \frac{1}{96}\Bigg\lbrace 
	160\, \partial_{k}C_{5,k} + 64\, \partial_{k}C_{7,k} + \mathcal{V}^{-1}
	\frac{\partial}{\partial p_{2,\mu}}\frac{\partial}{\partial p_{4,\mu}}
	\frac{\partial}{\partial p_{2,\nu}}\frac{\partial}{\partial p_{4,\nu}}
	\bigg|_{p_{2}\, =\, p_{4}\, =\, 0}
	\frac{\delta^{4}\partial_{k}\Gamma_{k}}{\delta\pi^{1}(-p_{2}-p_{4})
	\delta\pi^{2}(p_{2})\delta\pi^{1}(0)\delta\pi^{2}(p_{4})} \nonumber\\
	& & \qquad\ - \frac{5}{24} \mathcal{V}^{-1}
	\frac{\partial}{\partial p_{\mu}}
	\frac{\partial}{\partial p_{\mu}}
	\frac{\partial}{\partial p_{\nu}}\frac{\partial}{\partial p_{\nu}}
	\bigg|_{p\, =\, 0}
	\frac{\delta^{4}\partial_{k}\Gamma_{k}}{\delta\pi^{1}(-p)
	\delta\pi^{2}(0)\delta\pi^{1}(p)\delta\pi^{2}(0)}
	\Bigg\rbrace , \label{eq:c8} 
\end{IEEEeqnarray}
\end{widetext}
where the propagator $\bigl(\Gamma_{\! k}^{(2)} + R_{k}\bigr)^{-1}$ is represented 
by a (solid or dashed) line (depending on its bosonic or fermionic character) and 
the supertrace obviously translates into a closed loop. The regulator insertion
$\partial_{k}R_{k}$ is denoted by a ``crossed'' circle at the top of the loops. Moreover,
the functional derivatives generate external legs, such that the flow involves two-point 
and even four-point functions, and each diagram on the right-hand side includes all
possible permutations of them. These legs come with the momentum $p$ or $p_{i}$, 
$i \in \lbrace 1,2,3,4 \rbrace$, and the factor $\mathcal{V}$ is the (infinite) 
space-time volume. Five and six-point vertices stemming from the effective potential
are truncated. Finally, the different d.o.f.\ are depicted using the following colors: 
$\sigma\rightarrow\text{blue}$, $\vec{\pi}\rightarrow\text{red}$, $\pi^{4}\rightarrow
\text{orange}$, $\pi^{5}\rightarrow\text{violet}$, and $\psi\rightarrow\text{black}$;
this is further true for the regulator insertions.

From Eq.\ (\ref{eq:z2}) on, we omitted the diagrams for reasons of clarity [for all 
pion four-point functions, the diagrams are the same as in Eq.\ (\ref{eq:c2})]. Note 
also that we have only given one single equation for the pNGB wave-function renormalization 
$Z_{k}^{\boldsymbol{\pi}}$ (i.e., the one of the ``ordinary'' pions), since all of them 
are degenerate due to the symmetry of the QMDM in vacuum. The analytic expressions 
of the flow equations are obtained by using the \texttt{Mathematica} packages
\texttt{FeynCalc} \cite{Mertig:1990an, Shtabovenko:2016sxi, Shtabovenko:2020gxv}, 
\texttt{DoFun} \cite{Huber:2011qr, Huber:2019dkb}, and \texttt{FormTracer} 
\cite{Cyrol:2016zqb}. 

To solve the FRG flow equations, the effective potential $U_{k}$ is approximated by 
the Taylor polynomial
\begin{equation}
	U_{k}\!\left(\rho\right) = \sum_{n\, =\, 1}^{N}
	\frac{\alpha_{n,k}}{n!} \left(\rho - \chi\right)^{n},
\end{equation}
with the $k$-independent expansion point $\chi$. The choice of $N = 6$ has proven to 
be numerically stable \cite{Pawlowski:2014zaa}. The Taylor coefficients $\alpha_{n,k}$,
$n = 1,\ldots , 6$, are computed from Eq.\ (\ref{eq:potloops}) as
\begin{equation}
	\partial_{k}\alpha_{n,k} = \left.\frac{\partial^{n}}{\partial\rho^{n}}
	\right|_{\chi} \partial_{k}U_{k}.
\end{equation}
The second coefficient $\alpha_{2,k}$ coincides (up to a factor of minus one-half) 
with the momentum-independent quartic interaction as mentioned in Sec.\ \ref{sec:methods}
and Sec.\ \ref{sec:results},
\begin{equation}
	C_{1,k} = - \frac{\alpha_{2,k}}{2}. \label{eq:c1}
\end{equation}
All other variables are also evaluated at the point $\rho = \chi$, which is chosen
to be slightly larger than the IR minimum of $U_{k}$. The initialization values of the FRG flow 
for the two different UV-cutoff scenarios discussed in Sec.\ \ref{sec:results} are 
summarized in Table \ref{tab:UVpars}. We emphasize once more that the higher-derivative 
couplings $C_{2,k}$, $Z_{2,k}$, and $C_{i,k}$, $i = 3,\ldots , 8$, start at zero in 
the UV and are therefore solely generated from quantum fluctuations during the FRG
integration process.
\begin{table}[b!]
	\caption{\label{tab:UVpars}UV parameters.}
	\begin{ruledtabular}
		\begin{tabular}{l c c}
		\textbf{Parameter}
		& $\boldsymbol{\Lambda_{\mathrm{UV}} = 500}\ \mathbf{MeV}$ 
		& $\boldsymbol{\Lambda_{\mathrm{UV}} = 1}\ \mathbf{GeV}$ \\
		\colrule\\[-0.3cm]
		$\chi$ & $\left(53.03\ \mathrm{MeV}\right)^{2}$ 
		& $\left(32.30\ \mathrm{MeV}\right)^{2}$ \\[0.1cm]
		$\alpha_{1,\Lambda_{\mathrm{UV}}}$ & $\left(514\ \mathrm{MeV}\right)^{2}$ 
		& $\left(1080\ \mathrm{MeV}\right)^{2}$ \\[0.1cm]
		$\alpha_{2,\Lambda_{\mathrm{UV}}}$ & 1 & 243 \\[0.1cm]
		$\alpha_{n,\Lambda_{\mathrm{UV}}}\ \lbrace n > 2 \rbrace$ & 0 & 0 \\[0.1cm] 
		$h_{\mathrm{ESB}}$ & $2.155 \times 10^{6}\ \mathrm{MeV}^{3}$ 
		& $3.635 \times 10^{6}\ \mathrm{MeV}^{3}$ \\[0.1cm]
		$y_{\Lambda_{\mathrm{UV}}}$ & 4.5 & 5.0 \\[0.1cm]
		$Z_{\Lambda_{\mathrm{UV}}}^{\boldsymbol{\pi}, \sigma, \psi}$ & 1 & 1 \\[0.1cm]
		$C_{i,\Lambda_{\mathrm{UV}}},\;\! i \in \lbrace 2,\ldots , 8 \rbrace$ 
		& 0 & 0 \\[0.1cm]
		$Z_{2,\Lambda_{\mathrm{UV}}}$ & 0 & 0 \\
		\end{tabular}
	\end{ruledtabular}
\end{table}

For the numerical integration, we employ exponential-type regulators,
\begin{equation}
	\begin{aligned}
	R_{k}^{\boldsymbol{\pi}}\!\left(q^{2}\right) = &\ Z_{k}^{\boldsymbol{\pi}} q^{2} 
	r\!\left(\frac{q^{2}}{k^{2}}\right) \! , \\[0.1cm]
	R_{k}^{\sigma}\!\left(q^{2}\right) = &\ Z_{k}^{\sigma} q^{2} 
	r\!\left(\frac{q^{2}}{k^{2}}\right) \! , \\[0.1cm]
	R_{k}^{\psi}(q) = &\ -i Z_{k}^{\psi} \gamma_{\mu}q_{\mu} 
	r\!\left(\frac{q^{2}}{k^{2}}\right) \! ,
	\end{aligned} \label{eq:reg}
\end{equation}
with the ``shape function''
\begin{equation}
	r(x) = \frac{1}{\exp(x) - 1}.
\end{equation}
As functions of the FRG scale $k$ and the momentum $q$, the regulators (\ref{eq:reg})
obey the limits
\begin{equation}
	\begin{aligned}
	R_{k} \rightarrow &\ 0 & 
	\text{for}\ k \rightarrow &\ 0, \\[0.1cm]
	R_{k} \rightarrow &\ \infty &
	\text{for}\ k \rightarrow &\ \Lambda_{\mathrm{UV}}
	\rightarrow \infty, \\[0.1cm]
	R_{k} > &\ 0 & 
	\text{for}\ q \rightarrow &\ 0, \\[0.1cm]
	R_{k} \rightarrow &\ 0 & 
	\text{for}\ q \rightarrow &\ \infty.
	\end{aligned}
\end{equation}
\begin{figure*}[t!]
	\centering
		\includegraphics{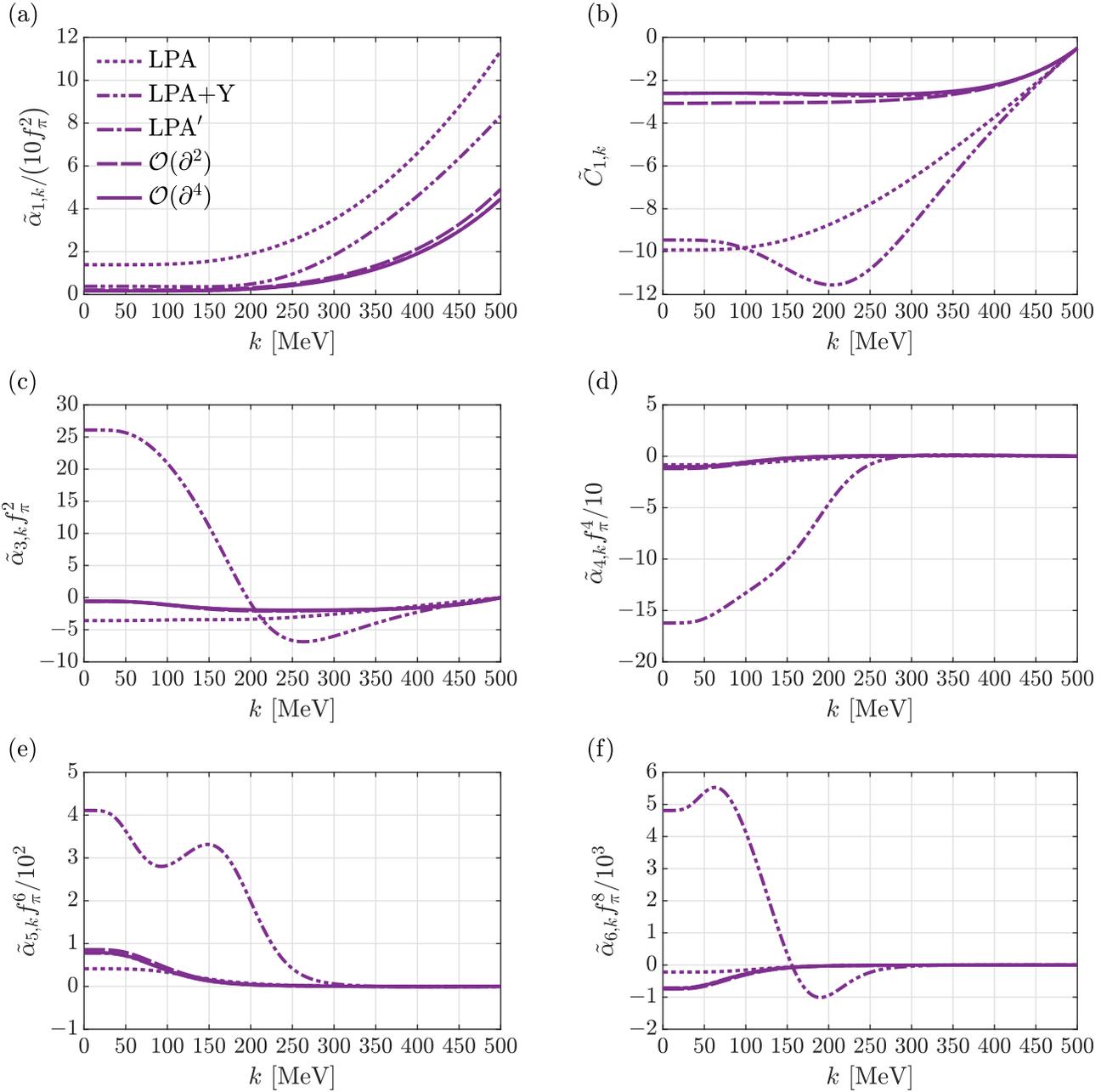}
	\caption{FRG flows of the (renormalized) Taylor coefficients ($\Lambda_{\mathrm{UV}} = 
	500\ \mathrm{MeV}$). The legend of subfigure (a) applies to all panels.}
	\label{fig:taylor_500}
\end{figure*}

\begin{figure*}[t!]
	\centering
		\includegraphics{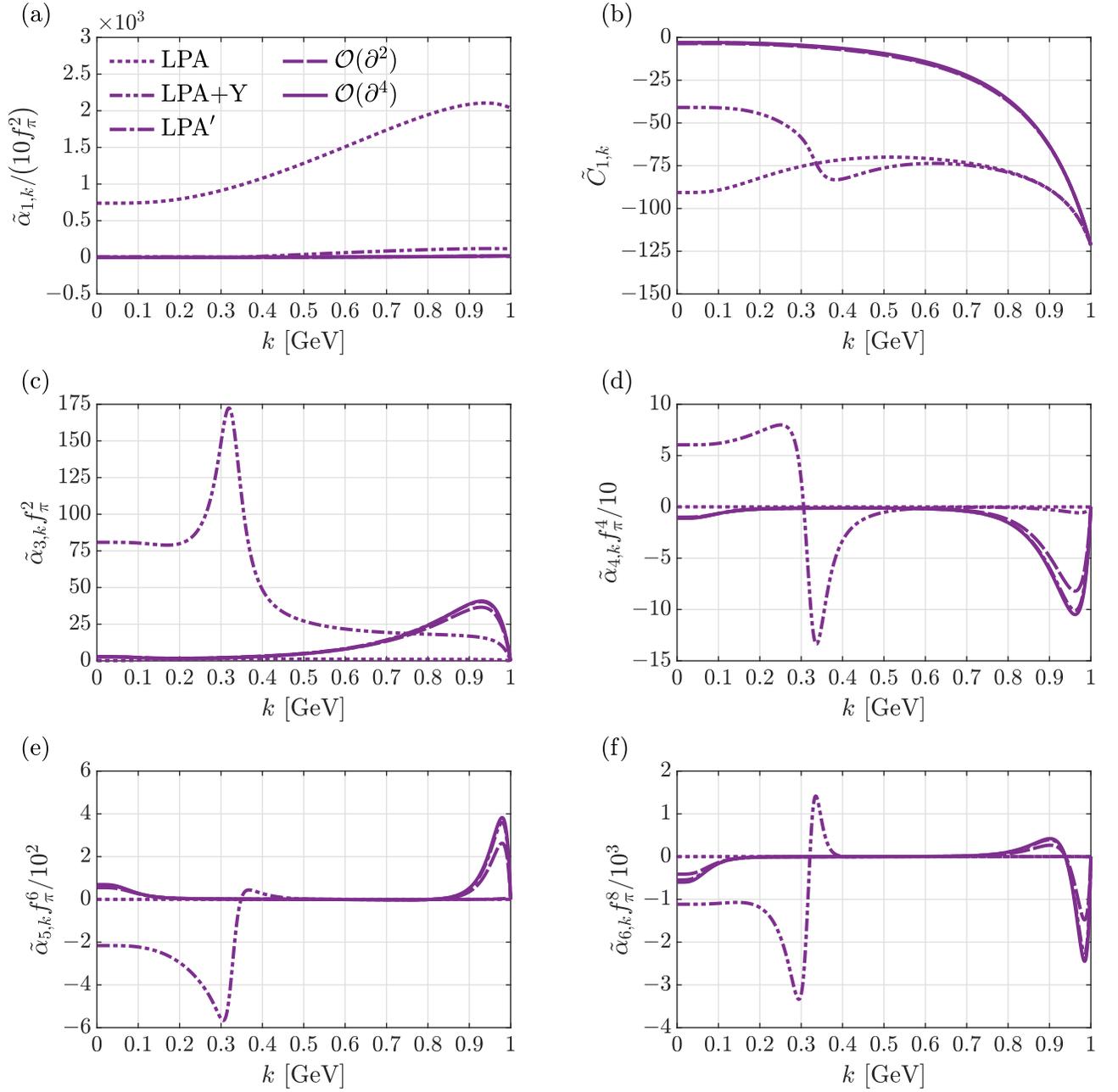}
	\caption{FRG flows of the (renormalized) Taylor coefficients ($\Lambda_{\mathrm{UV}} = 
	1\ \mathrm{GeV}$). The legend of subfigure (a) applies to all panels.}
	\label{fig:taylor_1000}
\end{figure*}

For the sake of completeness, we discuss the scaling behavior of the (renormalized) 
Taylor coefficients, $\tilde{\alpha}_{n,k} \equiv \alpha_{n,k}/(Z_{k}^{\boldsymbol{\pi}})^{n}$, 
$n = 1, \ldots , 6$, the Yukawa coupling $\tilde{y}_{k}$, and the wave-function renormalization 
factors $Z_{k}^{\boldsymbol{\pi}}$, $Z_{k}^{\sigma}$, and $\smash{Z_{k}^{\psi}}$ in this Appendix:
The Taylor coefficients (in dimensionless form) are plotted in Fig.\ \ref{fig:taylor_500} 
($\Lambda_{\mathrm{UV}} = 500\ \mathrm{MeV}$) and Fig.\ \ref{fig:taylor_1000}
($\Lambda_{\mathrm{UV}} = 1\ \mathrm{GeV}$). These flows determine the masses and the
decay constants, which are shown in Fig.\ \ref{fig:masses} in Sec.\ \ref{sec:results}. They exhibit 
strong corrections from the LPA and LPA+Y truncations towards the ``higher'' ones, as 
it has been documented above. However, the $\text{LPA}^{\prime}$ approximation is already 
close to the truncations including higher-derivative interactions, 
$\mathcal{O}\!\left(\partial^{2}\right)$ and $\mathcal{O}\!\left(\partial^{4}\right)$.
The values of the (renormalized) coefficients are quoted in Table \ref{tab:IRpars}
in Sec.\ \ref{sec:results}.
 
\begin{figure*}[t!]
	\centering
		\includegraphics{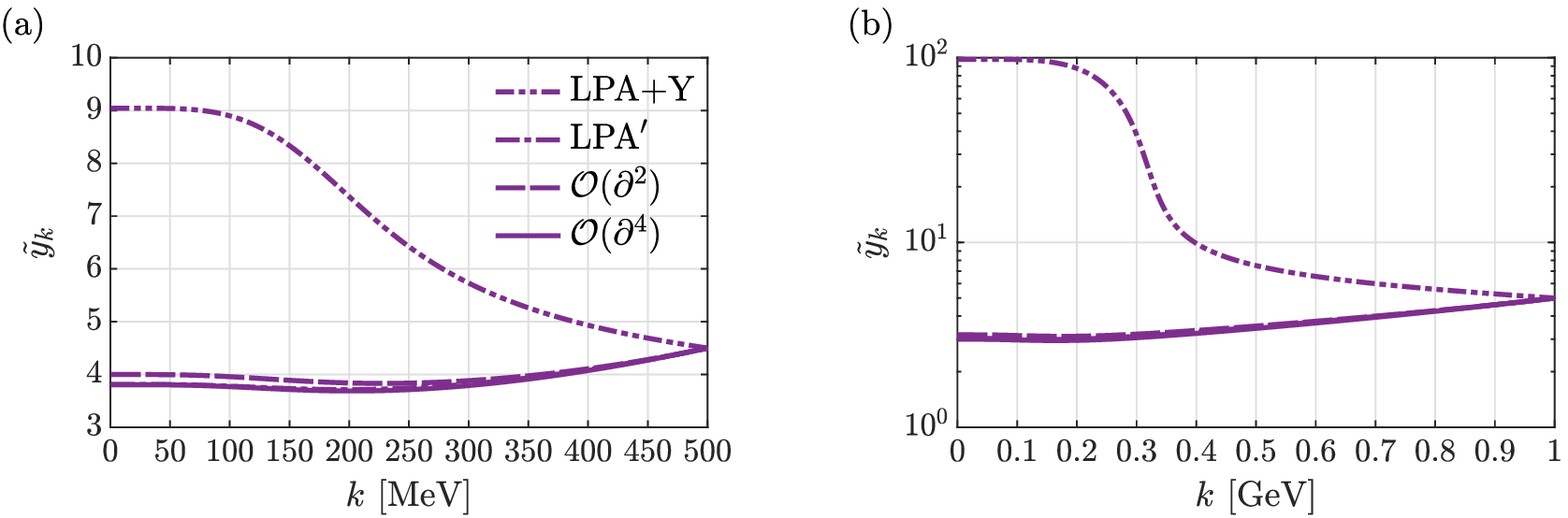}
	\caption{FRG flow of the (renormalized) Yukawa coupling $\tilde{y}_{k}$.
	The legend in subfigure (a) is also valid for subfigure (b).}
	\label{fig:yukawa}
\end{figure*}

Similarly to Fig.\ \ref{fig:masses} in Sec.\ \ref{sec:results} as well as Fig.\
\ref{fig:taylor_500} and Fig.\ \ref{fig:taylor_1000}, Fig.\ \ref{fig:yukawa} 
underlines the importance of the corrections from the wave-function renormalization, 
as compared to the (crudest) truncation LPA+Y. Within the latter, the value of $y_{k}$ 
increases in total to its IR-limit value, while it decreases in ``higher'' truncations, 
cf.\ Fig.\ \ref{fig:yukawa}(a) and Table \ref{tab:IRpars}. This difference is even 
more drastic for the scenario with a larger UV cutoff shown in Fig.\ \ref{fig:yukawa}(b)
and we end up with a quite ``mild'' flow of the parameter $\tilde{y}_{k}$, see also
Ref.\ \cite{Khan:2015puu} for very similar results.

\begin{figure*}[t!]
	\centering
		\includegraphics{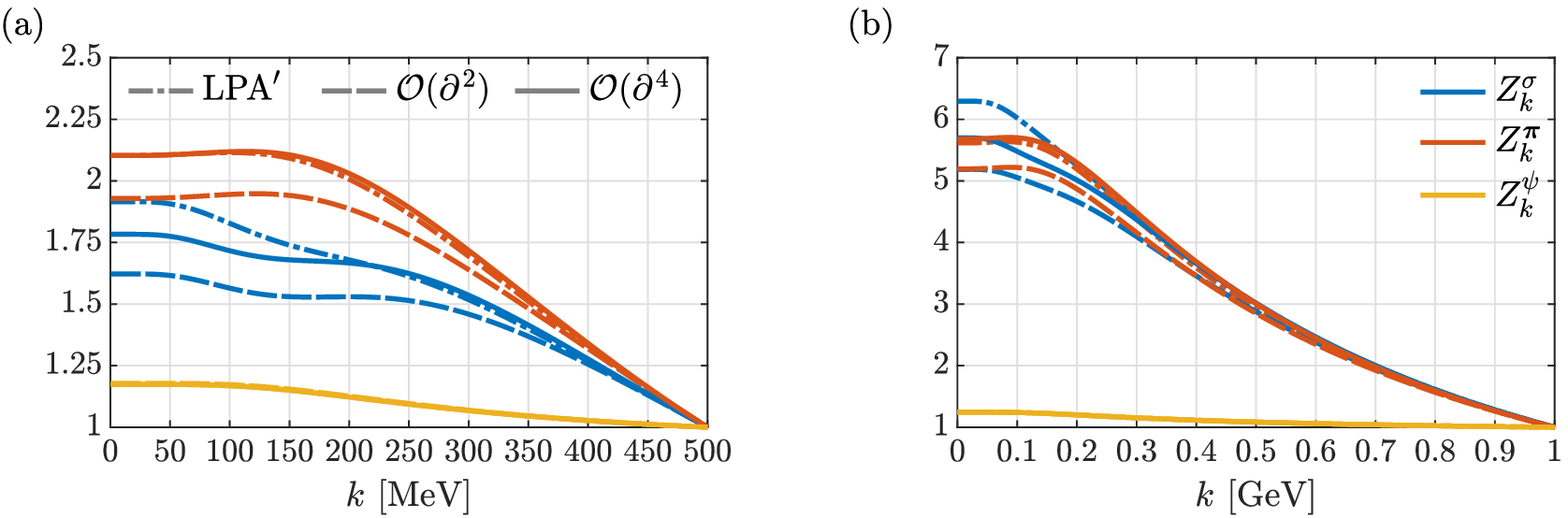}
	\caption{Scale evolution of the bosonic and fermionic wave-function renormalization 
	factors $Z_{k}^{\boldsymbol{\pi}}$, $Z_{k}^{\sigma}$, and $Z_{k}^{\psi}$. The
	legends in subfigures (a) and (b) apply to all panels.}
	\label{fig:wave}
\end{figure*}

Starting from their initialization at one, the bosonic wave-function renormalization 
factors $Z_{k}^{\boldsymbol{\pi}}$ and $Z_{k}^{\sigma}$ increase during the flow.
This behavior is shown in Fig.\ \ref{fig:wave}. Especially, the extra growth of 
the $Z_{k}^{\sigma}$-parameter below roughly 150 MeV is ascribed to the pure pion 
loops exclusively occurring in Eq.\ (\ref{eq:zs2}) [and not in Eq.\ (\ref{eq:zp2})]. 
This scale matches the decoupling of the quark fields from the FRG flow, as it is 
depicted in Fig.\ \ref{fig:cnonlinear}. Interestingly, the findings for the wave-function
renormalization in the $\text{LPA}^{\prime}$ approximation are closest to the ``full'' 
result of $\mathcal{O}\!\left(\partial^{4}\right)$. This phenomenon is inherited 
to many other couplings and coefficients, see once again Table \ref{tab:IRpars}. 
Finally, there is basically no significant difference between the various truncations for 
the scale evolution of the fermionic wave-function renormalization $Z_{k}^{\psi}$.

\bibliography{bib_LECs}

\end{document}